\documentclass[11pt]{article}

\RequirePackage{tgtermes}
\RequirePackage{newtxtext}
\RequirePackage{newtxmath}
\RequirePackage{bm}
\RequirePackage{endnotes}

\usepackage{mathtools, enumitem, bbm, xcolor, multirow, longtable}

\definecolor{CBblue}{RGB}{86,180,233}   
\definecolor{CBgreen}{RGB}{0,158,115}   
\definecolor{CBorange}{RGB}{213,94,0}   

\usepackage{amssymb,amsmath,amsthm, amsfonts}
\DeclareMathOperator*{\argmin}{arg\,min}
\DeclareMathOperator*{\argmax}{arg\,max}

\usepackage{fullpage}

\newtheoremstyle{myplain}
  {\topsep}   
  {\topsep}   
  {\itshape}  
  {}          
  {\bfseries} 
  {.}         
  {0.5em}     
  {}          

\theoremstyle{myplain}
\newtheorem{remark}{Remark}
\newtheorem{conjecture}{Conjecture}
\newtheorem{lemma}{Lemma}
\newtheorem{claim}{Claim}
\newtheorem{theorem}{Theorem}
\newtheorem{example}{Example}
\newtheorem{corollary}{Corollary}

\usepackage[T1]{fontenc}
\usepackage[utf8]{inputenc}
\usepackage{authblk} 

\newcommand{\bqo}{\operatorname{BQO}}
\newcommand{\rbqo}{\operatorname{R-BQO}}
\newcommand{\qubo}{\operatorname{QUBO}}

\newcommand{\rqubo}{\operatorname{R-QUBO}}

\newcommand{\st}{\operatorname{s.t.}}
\newcommand{\dsum}{\displaystyle \sum}

\usepackage{hyperref}
\hypersetup{colorlinks,linkcolor={black},citecolor={black},urlcolor={black}}

\usepackage{xfrac}
\usepackage{algorithm}
\usepackage[noend]{algpseudocode}

\usepackage{tikz}
\usetikzlibrary{arrows.meta,patterns}
\tikzstyle{vertex}=[circle, draw, inner sep=0pt, minimum size=1.5em]
\tikzstyle{svertex}=[square, draw, inner sep=0pt, minimum size=1.5em]
\newcommand{\vertex}{\node[vertex]}

\usepackage{natbib}
 \bibpunct[, ]{(}{)}{,}{a}{}{,}%

\usepackage{rotating}
\usepackage{fancyvrb}
\usepackage{braket}
\usepackage{fix-cm}

\title{Characterizing QUBO Reformulations of the Max-$k$-Cut Problem for Quantum Computing}

\author[1,2]{Adrian Harkness}
\author[3]{Hamidreza Validi}
\author[1]{Ramin Fakhimi}
\author[4]{Illya V. Hicks}
\author[2]{Samuel Stein}
\author[1]{Tam\'as Terlaky}
\author[1]{Luis F. Zuluaga}

\affil[1]{Department of Industrial and Systems Engineering, Lehigh University, Bethlehem, PA, US}
\affil[2]{Physical and Computational Sciences, Pacific Northwest National Laboratory, Richland, WA, US}
\affil[3]{Department of Industrial, Manufacturing \& Systems Engineering, Texas Tech University, TX, US}
\affil[4]{Computational Applied Mathematics \& Operations Research, Rice University, TX, US}

\date{} 

\begin{document}
\maketitle

\begin{abstract}

Quantum computing offers significant potential for solving NP-hard combinatorial (optimization) problems that are beyond the reach of classical computers. One way to tap into this potential is by reformulating combinatorial problems as a quadratic unconstrained binary optimization (QUBO) problem. The solution of the QUBO reformulation can then be addressed using adiabatic quantum computing devices or appropriate quantum computing algorithms on gate-based quantum computing devices. In general, QUBO reformulations of combinatorial problems can be readily obtained by properly penalizing the violation of the problem's constraints in the original problem's objective. However, characterizing tight (i.e., minimal but sufficient) penalty coefficients for this purpose is important and non-trivial for enabling the solution of the resulting QUBO in current and near-term quantum computing devices. Along these lines, we 
present closed-form characterizations of tight penalty coefficients for two distinct QUBO reformulations of the max $k$-cut problem whose values depend on the (weighted) degree of the vertices of the graph defining the problem. 
These findings contribute to the ongoing effort to make quantum computing a viable tool for solving combinatorial problems at scale. We support our theoretical results with illustrative examples and simple numerical results.

\end{abstract}

\maketitle


\section{Introduction}\label{sec1}

Quantum computing has significant potential to address the solution of NP-hard combinatorial (optimization) problems at sizes that are intractable for classical computers within reasonable time scales~\citep[see, e.g.,][]{abbas2024challenges}. 
One way to tap into this potential is by reformulating combinatorial problems as a
{\em quadratic unconstrained binary optimization} (QUBO) problem; that is, as an optimization problem with quadratic objective and the sole constraint that its variables are binary (i.e., with values $\{-1,1\}$ or $\{0,1\}$)~\citep[see, e.g.,][]{punnen2022}. The QUBO reformulation can then be solved using  {\em adiabatic quantum computing}~\citep[e.g.,][]{albash2018adiabatic} on {\em adiabatic-based quantum} devices~\citep[see, e.g.,][]{willsch2022benchmarking}, or the {\em quantum approximation optimization algorithm} (QAOA)~\citep{farhi2014_quantum} on {\em gate-based quantum} devices~\citep[see, e.g.,][]{kwon2021gate}.

Unconstrained combinatorial problems such as the max-cut problem or the equivalent {\em Ising model}~\citep{cipra2000ising} have a natural QUBO formulation. For constrained combinatorial problems, QUBO reformulations can be obtained by penalizing constraint violations in the original problem's objective
using large enough penalty coefficients that ensure that the problem constraints are satisfied by any optimal solution of the QUBO problem. This is a classical technique~\citep{abello2001finding,pardalos2006continuous} that has attracted renewed interest due to the emergence of quantum computing~\citep[see, e.g.,][]{lucas2014_ising, nannicini2019_performance, stollenwerk2019quantum, date2021qubo, dunning2018works, du2025solving}. 
However, simple large enough penalty coefficients for this purpose tend to grow with $\mathcal{O}(n)$, where $n$ measures the size of the problem~\citep[see, e.g.,][]{nannicini2019_performance, du2025solving, cruz2019qubo}. Such overly large penalties can negatively affect the performance of quantum devices~\citep[see, e.g.,][]{garcia2022exact, verma2022penalty,brown2024copositive,quintero2021_characterization} and 
make the use of the penalization method not scalable. Thus, it is important to characterize the tightest possible penalty coefficients for a given combinatorial problem~\citep[see, e.g.,][]{quintero2021_characterization, guney2025qubo}. 
Although numerical heuristics have been proposed to estimate tight penalty coefficients~\citep[see, e.g.,][]{garcia2022exact, verma2022penalty}, having a closed-form characterization of the tightest possible penalties makes it possible to fine-tune the best penalty coefficient, in terms of numerical performance, for the combinatorial problem~\citep[see, e.g.,][Fig. 13]{brown2024copositive}. Also, numerical heuristics can provide penalty coefficients that are erroneous; that is, their value is too low and does not lead to a QUBO reformulation of the problem. Indeed, experiments performed by the authors that are omitted for brevity show that this can be the case for the default heuristic used by {\tt Qiskit}'s \texttt{QuadraticProgramToQubo} converter, which is a general-purpose heuristic that does not guarantee equivalence for the max $k$-cut problem~\citep{QiskitConvertersTutorial,QiskitMinimumEigenOptimizer,QiskitQuadProgToQubo}.

Along these lines, our main contribution is to derive an analytical characterization of the penalty coefficients associated with two QUBO reformulations of the {\em max $k$-cut} problem~\citep[see, e.g.,][]{frieze1997_improved}, a fundamental NP-hard constrained combinatorial problem in which the aim is to partition the vertices of a graph into $k$ subsets such that the weight of the edges whose endpoints belong to different vertex subsets is maximized. One reformulation is based on the natural binary quadratic model of max $k$-cut problem (Theorem~\ref{theorem: optimal maxKcut qubo formulation}; see also~\eqref{eq: maxKcut_qp}), and the other one on a straightforward variable-reduced version of the natural max $k$-cut problem formulation (Theorem~\ref{thm:rqubo}; see also~\eqref{eq:maxKcut_rqp}). In both cases, we obtain the tightest penalty coefficients for instances with nonnegative edge weights~(Corollaries~\ref{cor:qubopos} and~\ref{cor:rqubopos}). Unlike penalty coefficients previously used for this problem, which depend directly on the number of vertices or edges of the graph~\citep[see, e.g.,][after eq.~(9)]{fuchs2021_efficient}, our coefficients depend on vertex degrees.
This is key to scaling the QUBO reformulation solution approach on quantum devices to larger-scale instances of the max $k$-cut problem,
as most graphs encountered in practical applications --- from social and biological networks to communication graphs --- are sparse~\citep[see, e.g.,][]{casiraghi2025empirical}. 

Throughout, we illustrate our results with appropriate examples. Further, these examples allow us to conjecture (see Conjectures~\ref{conj:qubo} and~\ref{conjecture: optimal maxKcut rqubo formulation}) that the tightest penalty coefficients for instances of the max $k$-cut problem with negative weights are close to the proven valid lower bounds on the penalty coefficients we provide (see Theorems~\ref{theorem: optimal maxKcut qubo formulation} and~\ref{thm:rqubo}). Strong evidence for the correctness of these conjectures comes from the examples in Appendices~\ref{sec:appconj} and~\ref{sec:appconj2}, which show that smaller coefficients can fail.

\looseness -1
We also present some simple numerical results  (Section~\ref{sec:num}) in which the proposed QUBO reformulations 
are solved using QAOA on a quantum computing simulator to illustrate the type of fine-tuning analysis of the penalty coefficients that our results enable. All the data and code used to generate our numerical results are publicly available in a {\tt GitHub} repository (see Section~\ref{sec:num} for details). 

Although our numerical experiments are conducted using QAOA, our results are readily useful when using other quantum computing approaches to solve the max $k$-cut problem. In particular, they enable the use of adiabatic quantum computing and \emph{Ising machines}~\citep{mohseni2022ising}, which have demonstrated promising performance on QUBO reformulations of combinatorial problems of nontrivial scale~\citep[see, e.g.,][]{brown2024copositive,quintero2021_characterization}.  
Furthermore, our results allow for the use of the {\em quantum Hamiltonian descent} (QHD) algorithm~\citep{leng2023quantum}, as the binary constraints in the QUBO reformulations can be relaxed following the approach described in~\cite{rosenberg1972breves}.
  
It is important to note that, when solving constrained combinatorial problems on gate-based quantum devices, an advantageous alternative to the penalization approach described above is the \emph{constraint-preserving} approach. This approach has been investigated in several works~\citep[see, e.g.,][]{Hadfield2019_from,wang2020xy, fuchs2024lx}.
In this approach, rather than making infeasible solutions of the combinatorial problem suboptimal by penalizing the constraints, constraint-preserving {\em mixers} are designed to ensure that,  during the evolution of a modified version of QAOA (referred to as the
quantum alternating operator ansatz), only feasible solutions of the combinatorial problem are considered (see Section~\ref{sec:num} for additional details). This strategy, together with efficient binary encoding techniques, has been used for the max $k$-cut problem in~\cite{fuchs2021_efficient, fuchs2025encodings}. The two approaches involve different trade-offs: penalization increases the complexity of the {\em cost} Hamiltonian, whereas constraint-preserving mixers shift this complexity to the mixer and circuit depth, a trade-off we quantify in Appendix~\ref{app:resources}. As illustrated in Section~\ref{sec:num}, under the hardware-informed noisy simulation considered there, penalization in combination with constraint-preserving mixers may provide a complementary mechanism for balancing solution quality and feasibility under noise. Another alternative to penalization in the framework of using gate-based quantum computers is the approach proposed by~\cite{bucher2025penalty}.

\section{Preliminaries}\label{section: background}

The max $k$-cut problem aims to partition the vertex set $V$ of a graph $G=(V,E)$ into $k \ge 2$ subsets such that the weight of the {\em cut edges} (i.e., edges whose endpoints belong to different vertex subsets) is maximized and can be formulated as a binary quadratic optimization ($\bqo$) problem as follows.  Let $n \coloneqq |V|$ and $m \coloneqq |E|$ denote the number of vertices and edges, respectively, and $P \coloneqq \{1, \dots, k\}$. For every edge~$\{u,v\}$, let $w_{uv} \in \mathbb{R}$ be the weight of edge $\{u,v\}$. Further, for every vertex $v \in V$ and every $j \in P$, the binary variable $x_{vj}$ is one if vertex $v$ is assigned to set $j$ of the partition $P$ and zero otherwise.  
Then, the max $k$-cut problem can be formulated as
\begin{subequations}\label{eq: maxKcut_qp}
	\begin{align}
	&& \max_{x \in \{0,1\}^{n \times k}} &&& g(x)\coloneqq \sum_{\{u,v\} \in E} w_{uv}\Big(1 - \sum_{j \in P} x_{uj}x_{vj} \Big)\label{eq: maxKcut_qp_1}\\
      (\bqo) && \st &&&\sum_{j \in P} x_{vj} =1, & \qquad \forall v \in V\label{eq: maxKcut_qp_2} 
	\end{align}
\end{subequations}
Above, the objective function~\eqref{eq: maxKcut_qp_1} maximizes the total weight of the cut edges, and constraints~\eqref{eq: maxKcut_qp_2} imply that each vertex must be assigned to exactly one set of the partition~$P$.

\begin{remark}[Scope with respect to $k$]
\label{rem:scope-k}
Unless stated otherwise, throughout the paper we consider the case $k\ge 3$. The case $k=2$ coincides with the max-cut problem, which already admits a natural QUBO formulation.
\end{remark}

Now we introduce a reduced $\bqo$ ($\rbqo$) formulation of the $\bqo$ formulation above where the number of variables of the problem is reduced by $n$. As mentioned in the introduction, this reduction is very relevant when considering the use of current or near-term quantum devices to solve the max $k$-cut problem. The formulation follows by taking advantage of the fact that setting $k-1$ of the partition sets is enough to define all the partition sets. That is, for any feasible solution $x \in \{0,1\}^{n\times k}$ of the $\bqo$ formulation, it follows that for any $v \in V$,
\begin{equation}
\label{eq:transform}
    x_{vk} = 1 - \sum_{j \in P \setminus \{k\}} x_{vj}.
\end{equation}
Thus, the $\bqo$ formulation is equivalent to
\begin{subequations}\label{eq:maxKcut_rqp}
	\begin{align}
	\max_{x \in \{0,1\}^{n \times (k -1)}} &\sum_{\{u,v\} \in E} w_{uv} \left( 1- \sum_{j \in P\setminus \{k\}} x_{uj} x_{vj} - \left(1 -\sum_{j \in P\setminus \{k\}} x_{uj}\right) \left(1 - \sum_{j \in P\setminus \{k\}} x_{vj} \right) \right) \label{eq:maxKcut_rqp_objective}\\
	  (\rbqo) \qquad\st \quad &\sum_{j \in P\setminus \{k\}} x_{vj} \le 1,  \qquad \qquad  \forall v \in V  \label{eq:clusterCons_rqp}
	\end{align}
\end{subequations}

\begin{remark} {\em We say that a problem is a reformulation of another one provided that both their optimal objective values coincide and their sets of optimal solutions are isomorphic. For example, the $\rbqo$ and $\bqo$ problems above are clearly reformulations of each other.}
\end{remark}

In what follows, we characterize the QUBO reformulations of the max $k$-cut problem that can be obtained from its $\bqo$ and $\rbqo$ formulations.

\section{QUBO reformulations} \label{section: The unconstrained binary optimization}

QUBO reformulations of the max $k$-cut problem can be obtained by appropriately penalizing the problem constraints into the objective function of the $\bqo$ and $\rbqo$ formulations of the problem.

For this purpose, let $E^- \coloneqq \big\{\{u,v\} \in E \ | \ w_{uv} < 0\big\}$ and $E^+ \coloneqq \big\{\{u,v\} \in E \ | \ w_{uv} > 0\big\}$. Also, 
for every vertex $v \in V$, let $N_G(v)$ be the set of the neighbors of vertex $v$ in graph $G(V, E)$ and 
\begin{align}
    N^+_G(v) \coloneqq \Big\{u \in N_G(v) \ \big| \ w_{uv} > 0 \Big\}, \quad & \quad N^-_G(v) \coloneqq \Big\{u \in N_G(v) \ \big| \ w_{uv} < 0 \Big\}, \nonumber\\
    d^+_v \coloneqq \sum_{u \in N^+_G(v)} w_{uv}, \quad & \quad d^-_v \coloneqq \sum_{u \in N^-_G(v)} w_{uv}. \label{eq: positive and negative weight degree}
\end{align}

Next, we characterize two $\qubo$ reformulations of the max $k$-cut problem associated with the $\bqo$ and $\rbqo$ formulations of the max $k$-cut problem presented in Section~\ref{section: background}.

\subsection{QUBO reformulation of BQO}
\label{sec:BQO_QUBO}

We first characterize a $\qubo$ reformulation of the max $k$-cut problem derived from its $\bqo$ formulation~\eqref{eq: maxKcut_qp}. 
That is, the $\qubo$ reformulation is obtained by penalizing the constraints~\eqref{eq: maxKcut_qp_2} of the $\bqo$ formulation into the objective function using a penalty coefficient vector $c \in \mathbb{R}_+^n$. Specifically, consider the problem
\begin{align} 
(\qubo) \quad \max_{x \in \{0,1\}^{n \times k}} q(x) &\coloneqq \sum_{\{u,v\} \in E} w_{uv} \Big(1 - \sum_{j \in P} x_{uj} x_{vj}\Big) - \sum_{v \in V} c_v \Big(\sum_{j \in P} x_{vj} - 1 \Big)^2. \label{eq:maxKcut_qubo}
\end{align}
Abusing notation, we will at times refer to problem~\eqref{eq:maxKcut_qubo} as just $\qubo$ for brevity.
An optimal solution of the $\qubo$ formulation~\eqref{eq:maxKcut_qubo} is not necessarily a feasible solution of the max $k$-cut problem.
That is, a solution of the $\qubo$ formulation~\eqref{eq:maxKcut_qubo} is not feasible for the max $k$-cut problem if some vertices are not assigned to any partition set or are assigned to more than one partition set.
However, because the optimal objective value of the max $k$-cut problem is bounded, one can always set a large enough penalty coefficient vector $c$ in the $\qubo$ formulation~\eqref{eq:maxKcut_qubo} to ensure that any optimal solution of this problem is a feasible solution of the max $k$-cut problem. For example, in~\cite{fuchs2021_efficient}, after eq. (9) in their paper, it is shown that a large enough penalty coefficient vector for this purpose, in the case of unweighted graphs (i.e., when $w_{uv} \in \{0,1\}$ for all $\{u,v\} \in E$), can be obtained by setting 
\begin{equation}
\label{eq:cnaive}
c_v = \max\left \{\frac{n}{k}, km \right \},
\end{equation}
for all $v \in V$. Notice that as the size of the instance of the max $k$-cut problem increases, the penalty coefficients in~\eqref{eq:cnaive} grow as $\mathcal{O}(m)$.
This is relevant in our setting as, among other things, the input length of the constants that current quantum computers can handle is very limited~\citep[see, e.g.,][]{vyskovcil2019embedding}, and the increasing magnitude of the penalty coefficients can substantially and negatively impact the performance of quantum computers~\citep[see, e.g.,][Sec. 4]{quintero2021_characterization}.
An interesting question is then to find the smallest such penalty coefficients.
Lemma~\ref{lemma: qubo min c coef} provides a lower bound for the penalty coefficient vector $c$ in the $\qubo$ formulation~\eqref{eq:maxKcut_qubo} such that the solution of the $\qubo$ is feasible for the max $k$-cut problem.

\begin{lemma}\label{lemma: qubo min c coef}
Let $G(V, E)$ be a graph with edge weights $w_{uv}$ for all $\{u,v\} \in E$. Let $c \in \mathbb{R}^n_+$ be a penalty coefficient vector
and $\hat{x} \in \{0,1\}^{n \times k}$ be an optimal solution of the $\qubo$ formulation~\eqref{eq:maxKcut_qubo}. 
If 
$c_v > \max\{\tfrac{d^+_v}{k}, -\tfrac{3}{2} d^-_v\}$ for every vertex $v \in V$, 
then $\hat{x}$ is a feasible solution of the max $k$-cut problem.
\end{lemma}

\begin{proof}
Let $q(x) = q_1(x) + q_2(x)$ with
	\begin{equation}\label{eq: qubo q1q2}
    \begin{array}{llll}
	    q_1(x) \coloneqq \sum_{\{u,v\} \in E} w_{uv} -  \sum_{\{u,v\} \in E} w_{uv}\sum_{j \in P} x_{uj} x_{vj}, 
	    &&&
	    q_2(x) \coloneqq -\sum_{v \in V} c_v \Big(\sum_{j \in P} x_{vj} - 1 \Big)^2,
        \end{array}
	\end{equation}
	and for all $v \in V$, let
         \[
         t_v \coloneqq \sum_{j \in P} \hat{x}_{vj},
         \]
    be the number of partition sets to which vertex $v$ is assigned in the optimal solution of the $\qubo$ formulation~\eqref{eq:maxKcut_qubo}.
    Note that from the $\bqo$ formulation of the max $k$-cut problem, to prove the result, it is enough to show that $t_v = 1$ for all vertices $v \in V$.

    \begin{claim}
    \label{at_most_one}
    For every vertex $v \in V$, $t_v \le 1$.
	\end{claim}

	Without loss of generality, assume that $N_G(v) \neq \varnothing$; that is, $d^+_v - d^-_v > 0$. We prove the claim by contradiction.
	Assume there is a vertex $v \in V$ such that $t_v \ge 2$. Let $i \in \{j \in P: \hat{x}_{vj} = 1\}$.
	Now define $\check{x}(i) \in \{0,1\}^{n \times k}$ as follows: ($i$) for every vertex $u \in V \setminus \{v\}$ and every $j \in P$, set $\check{x}_{uj}(i) \coloneqq \hat{x}_{uj}$ and ($ii$) 
	set $\check{x}_{vi}(i)= 0$ and $\check{x}_{vj}(i) \coloneqq \hat{x}_{vj}$ for every $j \in P \setminus \{i\}$. The notation $\check{x}(i)$ used here (as well as similar notation later on), is used to emphasize that this solution is constructed by modifying the elements of $\hat{x}$ associated with the partition index $i$. Clearly, $\check{x}(i)$ is feasible for the $\qubo$ formulation~\eqref{eq:maxKcut_qubo}.

     From~\eqref{eq: qubo q1q2}, it follows that
	\begin{align}
	\begin{split}
	\label{eq: qubo at most two 3}
	    q_1(\check{x}(i)) - q_1(\hat{x}) &= \sum_{u \in N_G(v)} \sum_{j \in P} w_{uv}\hat{x}_{uj}\hat{x}_{vj} -  \sum_{u \in N_G(v)} \sum_{j \in P}  w_{uv}\check{x}_{uj}(i)\check{x}_{vj}(i)\\
	    & = \sum_{u \in N^+_G(v)} w_{uv}\hat{x}_{ui}\hat{x}_{vi} + \sum_{u \in N^-_G(v)} w_{uv}\hat{x}_{ui}\hat{x}_{vi}  \ge 0 + d_v^-, 
	    \end{split}
	    \end{align}
	where the first equality holds by the definition of $q_1(\cdot)$ (see~\eqref{eq: qubo q1q2}), and the inequality holds since we have $ \sum_{u \in N^+_G(v)} w_{uv}\hat{x}_{ui}\hat{x}_{vi} \ge 0$.
    Thus, 
	\begin{align*}
	    q(\check{x}(i)) - q(\hat{x})  = (q_1(\check{x}(i)) - q_1(\hat{x}) )+ (q_2(\check{x}(i)) - q_2(\hat{x}))
	    &\ge  d_v^-  +   c_v \Big[(t_v - 1)^2 - (t_v - 2)^2 \Big] \\
	    & =  d_v^- +  c_v (2t_v - 3)\\
	    & > d_v^- +  \max\bigg\{\frac{d^+_v}{k}, -\frac{3d^-_v}{2} \bigg\} (2t_v - 3)\\
	    & \ge \begin{cases}
	    3d_v^- (\frac{11}{6} - t_v) > 0 & \text{if } d^-_v < 0, \\
	    \frac{d^+_v}{k}(2t_v - 3) > 0 & \text{if } d^-_v = 0.
	    \end{cases}
	\end{align*}
Here, the first inequality holds by~\eqref{eq: qubo at most two 3}.
The second strict inequality holds by the assumption of the lemma.
Finally, the last set of inequalities hold because $d^+_v - d^-_v > 0$ and $t_v \ge 2$ by assumption.  
This contradicts the fact that $\hat{x}$ is an optimal solution of the $\qubo$ formulation~\eqref{eq:maxKcut_qubo}.
Hence, $t_v \le 1$.

\begin{claim}\label{at_least_one}
    For every vertex $v \in V$, $t_v \ge 1$.
\end{claim}

    We prove the claim by contradiction. 
    Assume there is a vertex $v \in V$ such that $t_v = 0$. Let
    \begin{equation}
    \label{eq:idef}
    i \in \argmin_{ j \in P} \left \{\sum_{u \in N_G(v)}w_{uv} \hat{x}_{uj} \right \}.
    \end{equation}
    Now define $\check{x}(i) \in \{0,1\}^{n \times k}$ as follows: ($i$) for every vertex $u \in V \setminus \{v\}$ and every $j \in P$, set $\check{x}_{uj}(i) \coloneqq \hat{x}_{uj}$ and ($ii$) 
	set $\check{x}_{vi}(i)= 1$ and $\check{x}_{vj}(i) \coloneqq \hat{x}_{vj}$ for every $j \in P \setminus \{i\}$. Clearly, $\check{x}(i)$ is feasible for the $\qubo$ formulation~\eqref{eq:maxKcut_qubo}.

    From~\eqref{eq: qubo q1q2}, it follows that
    \[
    \begin{array}{ll}
    q_1(\check{x}(i)) - q_1(\hat{x}) & =  - \dsum_{u \in N_G(v)} w_{uv} \dsum_{j \in P} (\check{x}_{uj}(i)\check{x}_{vj}(i) - \hat{x}_{uj}\hat{x}_{vj})  = - \dsum_{u \in N_G(v)} w_{uv} \hat{x}_{ui},
    \end{array}
    \]
    and $q_2(\check{x}(i)) - q_2(\hat{x}) = c_v$. Thus
    \begin{equation}\label{eq: qubo decomposition I_0 and negative weight}
	    q(\check{x}(i)) - q(\hat{x}) =  c_v - \sum_{u \in N_G(v)} w_{uv} \hat{x}_{ui}.
\end{equation}
Then, consider the following two possible cases. 

\begin{enumerate}[label=(\roman*), wide, labelindent=0pt]
	\item If $\sum_{u \in N_G(v)} w_{uv} \hat{x}_{ui} \ge 0$,
    then
	\begin{align}\label{eq: qubo inequality I_0 and negative weight}
	    c_v >  \frac{d^+_v}{k} \geq \frac{1}{k} \sum_{u \in N_G^+(v)} w_{uv} \sum_{j \in P} \hat{x}_{uj} = \frac{1}{k} \sum_{j \in P} \sum_{u \in N_G^+(v)} w_{uv} \hat{x}_{uj}
	    & \ge \min_{j \in P}\bigg\{\sum_{u \in N_G^+(v)} w_{uv} \hat{x}_{uj}\bigg\} \nonumber \\ 
	    & \ge \min_{j \in P}\bigg\{\sum_{u \in N_G(v)} w_{uv} \hat{x}_{uj}\bigg\} \nonumber \\
	    & = \sum_{u \in N_G(v)} w_{uv} \hat{x}_{ui} .
	\end{align}
	Here, the first inequality holds by assumption. 
	The second inequality is satisfied due to the definition of $d^+_v$ (see ~\eqref{eq: positive and negative weight degree}) and Claim~\ref{at_most_one}.
    The third inequality holds as the minimum value of a set of numbers is less than or equal to their average.
	The last inequality holds because for every vertex $u \in N^-_{G}(v)$, we have $w_{uv} < 0$. 
	The last equality holds by the definition of $i$ (see~\eqref{eq:idef}). Thus, from~\eqref{eq: qubo decomposition I_0 and negative weight} and \eqref{eq: qubo inequality I_0 and negative weight}, it follows that $q(\check{x}(i)) - q(\hat{x}) > 0$.
	 \item If $\sum_{u \in N_G(v)} w_{uv} \hat{x}_{ui} < 0$, then it follows from~\eqref{eq: qubo decomposition I_0 and negative weight} that $q(\check{x}(i)) - q(\hat{x}) > 0$.   
	\end{enumerate}
Thus, in both cases, we have $q(\check{x}(i)) > q(\hat{x})$, which contradicts the optimality of $\hat{x}$ for the $\qubo$ formulation~\eqref{eq:maxKcut_qubo}.

By Claims~\ref{at_most_one} and~\ref{at_least_one}, it follows that $t_v = 1$ for every $v \in V$.
Therefore, $\hat{x}$ is feasible for the max $k$-cut problem.

\end{proof}

Using Lemma~\ref{lemma: qubo min c coef}, and fact that the $\qubo$ formulation~\eqref{eq:maxKcut_qubo} is a relaxation of the max $k$-cut problem, we can derive a lower bound on the penalty coefficients $c_v$, for all $v \in V$, such that the $\qubo$ formulation~\eqref{eq:maxKcut_qubo} is a reformulation of the max $k$-cut problem.

\begin{theorem}\label{theorem: optimal maxKcut qubo formulation}
Let $G(V, E)$ be a graph with edge weights $w_{uv}$ for all $\{u,v\} \in E$.
    Let $\hat{x} \in \{0,1\}^{n \times k}$ be an optimal solution of the $\qubo$ formulation~\eqref{eq:maxKcut_qubo} with~$c_v >  \max\big\{\frac{d^+_v}{k}, -\frac{3}{2} d^-_v \big\}$ for every vertex $v \in V$.
    Then, $\hat{x}$ is an optimal solution of the max $k$-cut problem.
\end{theorem}

\begin{proof}
Let $x^* \in \{0,1\}^{n \times k}$ be an optimal solution of the BQO formulation~\eqref{eq: maxKcut_qp} of the max $k$-cut problem. By Lemma~\ref{lemma: qubo min c coef}, $\hat{x}$ is feasible for the max $k$-cut problem, and since $x^*$ is optimal for it, it follows that $g(x^*) \geq g(\hat{x})$.
Furthermore, since $q_2(\hat{x}) = q_2(x^*) = 0$ and $x^*$ is a feasible solution of the $\qubo$ formulation, then $g(\hat{x}) = q(\hat{x}) \geq q(x^*) = g(x^*)$.
Thus, it follows that $g(x^*) = g(\hat{x})$.

\end{proof}

Consider the case of the unweighted max $k$-cut problem. In this case, note that the penalty coefficients in~\eqref{eq:cnaive} are proportional to the size of the graph. In contrast, the penalty coefficients in Theorem~\ref{theorem: optimal maxKcut qubo formulation} can be bounded by $c_v \le \frac{\Delta(G)}{k}$ for all $v \in V$, where $\Delta (G)$ is the maximum degree of the input graph $G$. This is relevant, because most graphs encountered in practical applications --- from social and biological networks to communication graphs --- are sparse~\citep[see, e.g.,][]{casiraghi2025empirical}. For example, when a (possibly approximate) scale-free assumption applies~\citep[see, e.g.,][for a nuanced discussion of scale-free structure in real-world data]{broido2019scale}, the maximum degree $\Delta(G)$ grows more slowly than the size of the graph; that is, $\Delta(G) \sim n^{1/(\gamma-1)}$, for $\gamma > 2$, where $\gamma$ is the parameter defining the power-law distribution of the vertex degrees.

	 Next, we show that for instances of the max $k$-cut problem with nonnegative edge weights, Theorem~\ref{theorem: optimal maxKcut qubo formulation} provides the tightest lower bound for the penalty coefficients such that the $\qubo$ formulation~\eqref{eq:maxKcut_qubo} is a reformulation of the max $k$-cut problem.

     \begin{remark}
\label{rem:graphs}
{\em In the figures throughout the article, we graphically illustrate solutions to different $\qubo$ formulations of the max $k$-cut problem as follows. 
A colored vertex means that it is assigned to a partition set with vertices of that color. In particular, a vertex with two colors is assigned to two partition sets. A white vertex indicates that the vertex is not assigned to any partition set.}
\end{remark}

\begin{corollary}\label{cor:qubopos}
The QUBO formulation~\eqref{eq:maxKcut_qubo} is a reformulation of the max $k$-cut problem for all graphs  $G(V, E)$ with nonnegative edge weights $w_{uv}$ for all $\{u,v\} \in E$ only if $c_v \ge  \max\{\tfrac{d^+_v}{k}, -\tfrac{3}{2} d^-_v\} = \tfrac{d^+_v}{k}$ for every vertex $v \in V$.
\end{corollary}

\begin{proof} 
From Theorem~\ref{theorem: optimal maxKcut qubo formulation}, it is enough to show a graph for which the QUBO formulation~\eqref{eq:maxKcut_qubo} is not a reformulation of the max $k$-cut problem when the penalty coefficients violate the given bound. For that purpose, for a given $k$, let $G(V,E)$ be a clique on the vertices $V = \{u_1, \dots, u_k\} \cup \{\tilde{u}\}$, with edge weights given by $w_{u_iu_j} = M$ for all $1\le i < j \le k$, and $w_{u_i\tilde{u}} = \alpha$ for all $i=1,\dots,k$, for some $\alpha > 0$, and $M > \alpha k$. This graph is illustrated for the case $k=3$ in Figure~\ref{fig:counter1}~(left).
The optimal solutions of the max $k$-cut problem on this instance are obtained by setting $P_i = \{u_i, \tilde{u}\}$ for some $i=1,\dots,k$ and $P_j = \{u_j\}$ for all $j\neq i$, $j=1,\dots,k$, with optimal objective ($k$-cut) value $z^* = \tfrac{1}{2}Mk(k-1) + \alpha(k-1)$. This is illustrated for the case $k=3$ in Figure~\ref{fig:counter1}~(center).
Note that the alternative solutions of the form $P_i = \{u_i\}$ for all $i\not \in \{i',j\}$, $i=1,\dots,k$, $P_{i'} = \{u_{i'}, u_j\}$, $P_{j} = \{\tilde{u}\}$, for any $1 \le i' < j \le k$ have objective value $z^* + \alpha - M < z^*$. Since all edge weights are positive, $d^+_{u_i} = (k-1)M+\alpha$ (summing $k-1$ edges of weight $M$ and one of weight $\alpha$) and $d^+_{\tilde{u}} = k\alpha$. Now consider the QUBO formulation~\eqref{eq:maxKcut_qubo} for $G(V, E)$ with penalty coefficients $c_{u_i} > \tfrac{d_{u_i}^+}{k} = \frac{(k-1)M+\alpha}{k}$ and $c_{\tilde{u}} < \tfrac{d_{\tilde{u}}^+}{k} = \alpha$ (i.e., with the penalty coefficient associated with vertex $\tilde{u}$ violating the lower bound $\tfrac{d_{\tilde{u}}^+}{k}$), and the feasible solution for this problem defined by setting $x_{u_ij}= 1$ for $j=i$ and $x_{u_ij}= 0$ for all $j \in \{1,\dots,k\} \setminus \{i\}$ for $i=1,\dots,k$, and $x_{\tilde{u}j} = 0$ for $j=1,\dots,k$, which represents setting $P_i = \{u_i\}$, $i=1,\dots,k$ and not assigning $\tilde{u}$ to any partition (an infeasible assignment for the max $k$-cut problem). 
This is illustrated for the case $k=3$ in Figure~\ref{fig:counter1}~(right).
\begin{figure}[ht!]
\centering
		\includegraphics[scale = 0.76]{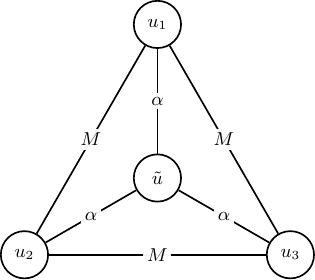}
		 \hspace{8mm}
		\includegraphics[scale = 0.76]{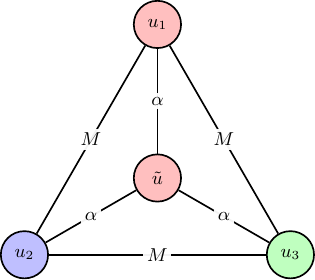}
		 \hspace{8mm}
		\includegraphics[scale = 0.76]{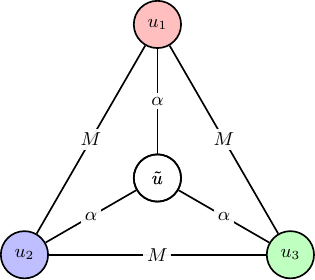}
\caption{Illustration of counterexample used in Corollary~\ref{cor:qubopos} for the case $k=3$. Left: Graph. Center: Optimal solution of max $k$-cut problem. Right: Feasible solution for QUBO formulation~\eqref{eq:maxKcut_qubo}. \label{fig:counter1}}
\end{figure}
For this solution (recall~\eqref{eq:maxKcut_qubo}), $q(x) =\tfrac{1}{2}Mk(k-1) + \alpha k - c_{\tilde{u}} = z^* + \alpha - c_{\tilde{u}} > z^*$. This shows that for the given graph, the optimal value of the QUBO formulation~\eqref{eq:maxKcut_qubo} with penalty coefficients violating the lower bounds provided in the result statement is greater than the optimal value of the max $k$-cut problem for this instance, which proves the desired result.
\end{proof}

However, the following example shows that Theorem~\ref{theorem: optimal maxKcut qubo formulation}
does not provide the tightest penalty coefficients 
when the graph $G(V, E)$ contains at least one edge with a negative weight.

\begin{remark}
\label{rem:opt}
{\em Throughout the examples provided in the article, the stated optimal objective values or solutions of the small-scale optimization problems are verified using the mixed-integer optimization solver {\tt Gurobi}~\citep{gurobi}.}
\end{remark}

\begin{example}\label{example: max_k_cut qubo counter example neq} 
{\em
Figure~\ref{fig:max_k_cut qubo counter example neq} illustrates an instance of the max 3-cut problem with optimal objective value of 5. 
The non-zero entries of an optimal solution $x^* \in \{0,1\}^{4 \times 3}$ of the max 3-cut problem are $x_{11}^* = x_{21}^* = 1$, $x_{32}^* = 1$, and $x_{43}^* = 1$. 
See Figure~\ref{fig:max_k_cut qubo counter example neq} (left) for an illustration.
The corresponding $\qubo$ formulation~\eqref{eq:maxKcut_qubo}  for this problem is as follows
	\begin{align}
    \label{eq:quboex2}
	    \max_{x \in \{0,1\}^{4 \times 3}} q(x) := 4 -  \sum_{\{u,v\} \in E} w_{uv}\left(x_{u1}x_{v1} + x_{u2}x_{v2} + x_{u3}x_{v3}\right) - \sum_{v \in V} c_v \left(x_{v1}+x_{v2}+x_{v3} - 1 \right)^2.
	\end{align}
    \begin{figure}[htb]
		\centering
    \begin{tikzpicture}[scale=0.15]

    \vertex(1) at (8,26) [fill=CBblue!30, label={[align=center,font=\fontsize{8}{0}\selectfont]}] {$1$};
    
    \vertex(2) at (8,10) [fill=CBblue!30, label={[align=center,font=\fontsize{8}{0}\selectfont]}] {$2$};
    
    \vertex(3) at (14,18) [fill=CBgreen!25, label={[align=center,font=\fontsize{8}{0}\selectfont]}] {$3$};
    
    \vertex(4) at (26,18) [fill=CBorange!30, label={[align=center,font=\fontsize{8}{0}\selectfont]}] {$4$};
    
    \draw (1) to [] node [midway, fill=white, font=\fontsize{8}{0}\selectfont] {$1$} (3);
    
    \draw (1) to [] node [midway, fill=white, font=\fontsize{8}{0}\selectfont] {$1$} (4);

    \draw (2) to [] node [midway, fill=white, font=\fontsize{8}{0}\selectfont] {$1$} (3);
    
    \draw (2) to [] node [midway, fill=white, font=\fontsize{8}{0}\selectfont] {$1$} (4);

    \draw (1) to [] node [midway, fill=white, font=\fontsize{8}{0}\selectfont] {$-1$} (2);

    \draw (3) to [] node [midway, fill=white, font=\fontsize{8}{0}\selectfont] {$1$} (4);

    \end{tikzpicture}\hspace{20mm}
	\begin{tikzpicture}[scale=0.15]

    \vertex(1) at (8,26) [fill=CBblue!30, label={[align=center,font=\fontsize{8}{0}\selectfont]}] {$1$};
    
    \vertex(2) at (8,10) [fill=CBblue!30, label={[align=center,font=\fontsize{8}{0}\selectfont]}] {$2$};
    
    \vertex(3) at (14,18) [fill=CBgreen!25, label={[align=center,font=\fontsize{8}{0}\selectfont]}] {$3$};
    
    \vertex(4) at (26,18) [fill=CBorange!30, label={[align=center,font=\fontsize{8}{0}\selectfont]}] {$4$};
    
    \draw (1) to [] node [midway, fill=white, font=\fontsize{8}{0}\selectfont] {$1$} (3);
    
    \draw (1) to [] node [midway, fill=white, font=\fontsize{8}{0}\selectfont] {$1$} (4);

    \draw (2) to [] node [midway, fill=white, font=\fontsize{8}{0}\selectfont] {$1$} (3);
    
    \draw (2) to [] node [midway, fill=white, font=\fontsize{8}{0}\selectfont] {$1$} (4);

    \draw (1) to [] node [midway, fill=white, font=\fontsize{8}{0}\selectfont] {$-1$} (2);

    \draw (3) to [] node [midway, fill=white, font=\fontsize{8}{0}\selectfont] {$1$} (4);
    \end{tikzpicture}

    \caption{Optimal solutions of the $\qubo$ formulation~\eqref{eq:maxKcut_qubo} for the max 3-cut problem with different penalty coefficients. With $\epsilon=10^{-3}$. Left: optimal solution with $c_1 = c_2 = 1.5 + \epsilon$ and $c_3 = c_4 = 1 + \epsilon$ that is optimal for the max $k$-cut problem. Right: optimal solution after changing $c_2$ from $1.5 + \epsilon$ to $1.5 - \epsilon$ that is still optimal for the max $k$-cut problem.}\label{fig:max_k_cut qubo counter example neq} 
	\end{figure}
	Let $\epsilon = 10^{-3}$. If we set the penalty coefficients as $c_1 = c_2 = -3/2 d^-_1 + \epsilon = -3/2 d^-_2 + \epsilon = 1.5 + \epsilon$ and $c_3 = c_4 = d^+_3/3 + \epsilon = d^+_4/3 + \epsilon = 1 + \epsilon$, satisfying the lower bound for the penalty coefficients in Theorem~\ref{theorem: optimal maxKcut qubo formulation},
    then the optimal solution of~\eqref{eq:quboex2} is $x^*$ with objective $q(x^*) = 5$ (see Figure~\ref{fig:max_k_cut qubo counter example neq} (left)).
	If we now change $c_2$ to $1.5 - \epsilon$, violating the lower bound for the penalty coefficients in Theorem~\ref{theorem: optimal maxKcut qubo formulation}, then the optimal solution of~\eqref{eq:quboex2} is still $x^*$ with objective $q(x^*) = 5$ (see Figure~\ref{fig:max_k_cut qubo counter example neq} (right)). Thus, slightly decreasing the penalty coefficient $c_2$ from $1.5 + \epsilon$ to $1.5-\epsilon$ 
    does not affect the optimality for the max $k$-cut problem of the optimal solutions obtained by the $\qubo$ formulation~\eqref{eq:maxKcut_qubo}. }
\end{example}

Although Theorem~\ref{theorem: optimal maxKcut qubo formulation} does not provide the tightest lower bound for the penalty coefficients in the case when the graph of interest contains negative edge weights, we conjecture that the general tightest lower bound is very close to the one provided in Theorem~\ref{theorem: optimal maxKcut qubo formulation}. Specifically, we conjecture that the tightest penalty coefficients are obtained by changing $-\frac{3}{2}d^-_v$ to $-\frac{1}{2}d^-_v$ for all $v \in V$ in Theorem~\ref{theorem: optimal maxKcut qubo formulation}. Positive evidence for this conjecture stems from unreported numerical experiments and Appendix~\ref{sec:appconj}.

\begin{conjecture}
\label{conj:qubo}
Let $G(V, E)$ be a graph with edge weights $w_{uv}$ for all $\{u,v\} \in E$.
    Let $\hat{x} \in \{0,1\}^{n \times k}$ be an optimal solution of the $\qubo$ formulation~\eqref{eq:maxKcut_qubo} with~$c_v >  \max\big\{ \frac{d^+_v}{k},  -\frac{d_v^-}{2
    } \big\}$ for every vertex $v \in V$.
    Then, $\hat{x}$ is an optimal solution of the max $k$-cut problem.
\end{conjecture}

\subsection{QUBO reformulation of R-BQO}
\label{sec:R-BQO_QUBO}

Similar to Section~\ref{sec:BQO_QUBO}, we now characterize a QUBO reformulation of the max $k$-cut problem derived, this time, from its R-BQO formulation~\eqref{eq:maxKcut_rqp}.
That is, the $\qubo$ reformulation is obtained by penalizing the constraints~\eqref{eq:clusterCons_rqp} of the $\rbqo$ formulation into the objective function using a penalty coefficient vector $c \in \mathbb{R}_+^n$. Specifically, let $\bar{P} \coloneqq P \setminus \{k\}$ and consider the problem
\begin{align}
     (\rqubo) \quad \max_{x \in \{0,1\}^{n \times {(k-1)}}} \bar{q}(x) \coloneqq 
     &  \sum_{\{u,v\} \in E} w_{uv} \Big[ 1- \sum_{j \in \bar{P}} x_{uj} x_{vj} - \Big(1 -\sum_{j \in \bar{P}} x_{uj}\Big) \Big(1 - \sum_{j \in \bar{P}} x_{vj} \Big) \Big]\nonumber \\
     &-\sum_{v \in V} c_v \sum_{\{i,j\} \in \binom{\bar{P}}{2} } x_{vi} x_{vj}. \label{eq: maxKcut_rqubo}
\end{align}
Analogous to the discussion at the beginning of Section~\ref{sec:BQO_QUBO}, one can always set a large enough penalty coefficient vector $c$ in the $\rqubo$ formulation~\eqref{eq: maxKcut_rqubo} to ensure that any optimal solution of this problem is a feasible solution of the max $k$-cut problem. 
However, an interesting question is then to find the smallest such penalty coefficients.
Lemma~\ref{lem:rqubo} provides a lower bound for the penalty coefficient vector $c$ in the $\rqubo$ formulation~\eqref{eq: maxKcut_rqubo} such that the solution of the $\rqubo$ is feasible for the max $k$-cut problem.

\begin{lemma}\label{lem:rqubo}
Let $G(V, E)$ be a graph with edge weights $w_{uv}$ for all $\{u,v\} \in E$. Let $c \in \mathbb{R}^n_+$ be a penalty coefficient vector and $\hat{x} \in \{0,1\}^{n \times (k-1)}$ be an optimal solution of $\rqubo$~\eqref{eq: maxKcut_rqubo}. If  
	$c_v > d_v^+ - 2d_v^-$ for every vertex $v\in V$,
	then $\hat{x}$ is a feasible solution of the max $k$-cut problem.
\end{lemma}

\begin{proof}
The proof is provided in Appendix~\ref{app:lemma2proof}.

\end{proof}

\begin{theorem}
\label{thm:rqubo}
Let $G(V, E)$ be a graph with edge weights $w_{uv}$ for all $\{u,v\} \in E$.
Let $\hat{x} \in \{0,1\}^{n \times (k-1)}$ be an optimal solution of $\rqubo$ with penalty coefficients $c \in \mathbb{R}^n_+$. If $c_v > d_v^+ - 2d_v^-$ for every $v\in V$, then $\hat{x}$ is an optimal solution of the max $k$-cut problem.
\end{theorem}

\begin{proof}
The proof is analogous to the proof of Theorem~\ref{theorem: optimal maxKcut qubo formulation}, but now using Lemma~\ref{lem:rqubo}. 
\end{proof}

Next, we show that for instances of the max $k$-cut problem with nonnegative edge weights, Theorem~\ref{thm:rqubo} provides the tightest lower bound for the penalty coefficients such that the $\rqubo$ is a reformulation of the max $k$-cut problem.

\begin{corollary}\label{cor:rqubopos}
The $\rqubo$ formulation~\eqref{eq: maxKcut_rqubo} is a reformulation of the max $k$-cut problem for all graphs $G(V, E)$ with nonnegative edge weights $w_{uv}$ for all $\{u,v\} \in E$ only if $c_v \ge d_v^+ - 2d_v^- = d_v^+$ for every vertex $v \in V$.
\end{corollary}

\begin{proof} 
From Theorem~\ref{thm:rqubo}, it is enough to show a graph for which the R-QUBO formulation~\eqref{eq: maxKcut_rqubo} is not a reformulation of the max $k$-cut problem when the penalty coefficients violate the given bound. For that purpose, for a given $k$, let $G(V,E)$ be a star with vertex set $V = \{u_1,\dots,u_k\} \cup \{\tilde{u}\}$ with vertex $\tilde{u}$ at the star's center and arbitrary nonnegative edge weights; that is, $E = \{u_1,\dots,u_k\} \times \{\tilde{u}\}$, and $w_{u_i \tilde{u}} > 0$ for all $i=1,\dots,k$. This graph is illustrated for the case $k=3$ in Figure~\ref{fig:counter2}~(left).
An optimal solution of the max $k$-cut problem on this instance is obtained by setting $P_1 = \{\tilde{u}\}$, $P_2 = \{u_1,\dots,u_k\}$, and $P_j = \emptyset$ for $j=3,\dots,k$, with optimal objective ($k$-cut) value $z^* = \sum_{i=1}^k w_{u_i \tilde{u}}$. This is illustrated for the case $k=3$ in Figure~\ref{fig:counter2}~(center). Now consider the R-QUBO formulation~\eqref{eq: maxKcut_rqubo} for $G(V, E)$ with penalty coefficients $c_{u_i} > d_{u_i}^+ = w_{u_i \tilde{u}}$ and $c_{\tilde{u}} < d_{\tilde{u}}^+ = \sum_{i=1}^k w_{u_i \tilde{u}} = z^*$ (i.e., with the penalty coefficient associated with vertex $\tilde{u}$ violating the lower bound $d_{\tilde{u}}^+$), and the feasible solution for this problem defined by setting $x_{u_ij}= 0$ for all $j \in \{1,\dots,k-1\}$ for $i=1,\dots,k$, and $x_{\tilde{u}1} = 1$, $x_{\tilde{u}2} = 1$, and $x_{\tilde{u}j} = 0$ for $j \in \{3,\dots,k-1\}$, which represents setting $P_1 = P_2 = \{\tilde{u}\}$, $P_k = \{u_1,\dots,u_k\}$, and $P_j = \emptyset$ for $j=3,\dots,k-1$ (an infeasible assignment for the max $k$-cut problem). 
This is illustrated for the case $k=3$ in Figure~\ref{fig:counter2}~(right).
For this solution (recall~\eqref{eq: maxKcut_rqubo}), $\bar{q}(x) =2\sum_{i=1}^k w_{u_i \tilde{u}} - c_{\tilde{u}} = 2z^* - c_{\tilde{u}} > z^*$. 
\begin{figure}[ht!]
\centering
		\includegraphics[scale = 0.76]{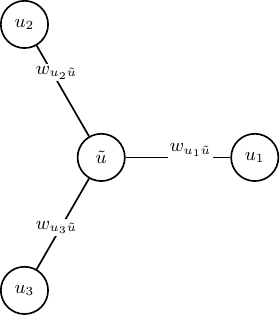}
		 \hspace{8mm}
		\includegraphics[scale = 0.76]{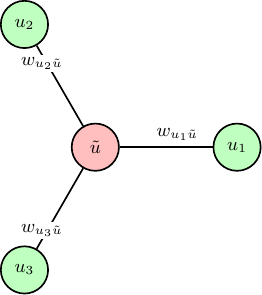}
		 \hspace{8mm}
		\includegraphics[scale = 0.76]{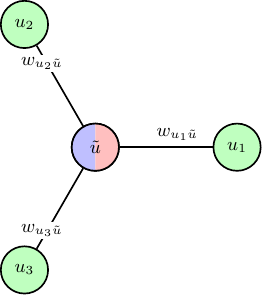}
\caption{Illustration of counterexample used in Corollary~\ref{cor:rqubopos} for the case $k=3$. Left: Graph. Center: Optimal solution of max $k$-cut problem. Right: Feasible solution for R-QUBO formulation~\eqref{eq: maxKcut_rqubo}. \label{fig:counter2}}
\end{figure}
This shows that for the given graph, the optimal value of the R-QUBO formulation~\eqref{eq: maxKcut_rqubo} with penalty coefficients violating the lower bounds provided in the result statement is greater than the optimal value of the max $k$-cut problem for this instance, which proves the desired result.
\end{proof}

Note that, for $k\ge 3$, the tightest penalty coefficients given in Corollary~\ref{cor:qubopos} for the QUBO reformulation of the max $k$-cut problem are therefore not valid for the R-QUBO formulation.
This fact highlights the need for the analysis in this section to obtain the tightest penalty coefficients for the R-QUBO formulation with a reduced number of binary variables, as discussed in the introduction, is very relevant for current and near-term quantum computers. On the other hand, the difference between the number of binary variables required by the QUBO formulation; that is, $nk$ and those required by the R-QUBO formulation; that is, $n(k-1)= nk - n$, should be relevant only for small values of $k$. Additionally, for a fixed graph with nonnegative edge weights, the tightest QUBO penalty coefficient scales as $1/k$ (so it decreases as $k$ grows), whereas the tightest R-QUBO penalty coefficient is independent of $k$. Thus, it can be argued that the R-QUBO formulation is particularly relevant for small values of $k$.

As the following example shows,  
Theorem~\ref{thm:rqubo} does not provide the tightest penalty coefficients when the graph $G(V,E)$ contains at least one edge with a negative weight.

\begin{example}\label{example:rqubo counter example1.5} 
{\em Figure~\ref{fig:max_k_cut rqubo counter example1.5} illustrates an instance of the max 3-cut problem with optimal objective value of 5. 
The non-zero entries of an optimal solution $x^* \in \{0,1\}^{5 \times 2}$ of the max 3-cut problem are $x_{31}^* = 1$, and $x_{42}^* = x_{52}^* = 1$. See Figure~\ref{fig:max_k_cut rqubo counter example1.5} (left) for an illustration.
The corresponding $\rqubo$ formulation~\eqref{eq: maxKcut_rqubo} for this problem is as follows
	\begin{equation}
    \label{eq:ex4.5}
	    \bar{q}(x) = 5 - \sum_{\{u,v\} \in E} w_{uv} \Big[ \sum_{j \in \{1,2\} } x_{uj}x_{vj} + \Big(1 -\sum_{j \in \{1,2\}} x_{uj}\Big)  \Big(1 -\sum_{j \in \{1,2\}} x_{vj}\Big) \Big] - \sum_{v \in V} c_v x_{v1} x_{v2}.
	\end{equation}
    \begin{figure}[htb]
		\centering
    \begin{tikzpicture}[scale=0.15]

     \vertex(1) at (0,18) [fill=CBorange!30, label={[align=center,font=\fontsize{8}{0}\selectfont]}] {$1$};
    
    \vertex(2) at (8,26) [fill=CBorange!30, label={[align=center,font=\fontsize{8}{0}\selectfont]}] {$2$};
    
    \vertex(3) at (8,10) [fill=CBblue!30, label={[align=center,font=\fontsize{8}{0}\selectfont]}] {$3$};
    
    \vertex(4) at (14,18) [fill=CBgreen!25, label={[align=center,font=\fontsize{8}{0}\selectfont]}] {$4$};
    
    \vertex(5) at (26,18) [fill=CBgreen!25, label={[align=center,font=\fontsize{8}{0}\selectfont]}] {$5$};

    \draw (1) to [] node [midway, fill=white, font=\fontsize{8}{0}\selectfont] {$-1$} (2);
    
    \draw (1) to [] node [midway, fill=white, font=\fontsize{8}{0}\selectfont] {$1$} (3);

    \draw (2) to [] node [midway, fill=white, font=\fontsize{8}{0}\selectfont] {$1$} (4);
    
    \draw (2) to [] node [midway, fill=white, font=\fontsize{8}{0}\selectfont] {$1$} (5);

    \draw (3) to [] node [midway, fill=white, font=\fontsize{8}{0}\selectfont] {$1$} (4);
    
    \draw (3) to [] node [midway, fill=white, font=\fontsize{8}{0}\selectfont] {$1$} (5);

    \end{tikzpicture}\hspace{10mm}
	\begin{tikzpicture}[scale=0.15]

    \vertex(1) at (0,18) [fill=CBorange!30, label={[align=center,font=\fontsize{8}{0}\selectfont]}] {$1$};
    
        \vertex(2) at (8,26) [fill=CBorange!30, label={[align=center,font=\fontsize{8}{0}\selectfont]}] {$2$};

    \vertex(3) at (8,10) [fill=CBblue!30, label={[align=center,font=\fontsize{8}{0}\selectfont]}] {$3$};
    
    \vertex(4) at (14,18) [fill=CBgreen!25, label={[align=center,font=\fontsize{8}{0}\selectfont]}] {$4$};
    
    \vertex(5) at (26,18) [fill=CBgreen!25, label={[align=center,font=\fontsize{8}{0}\selectfont]}] {$5$};

    \draw (1) to [] node [midway, fill=white, font=\fontsize{8}{0}\selectfont] {$-1$} (2);
    
    \draw (1) to [] node [midway, fill=white, font=\fontsize{8}{0}\selectfont] {$1$} (3);

    \draw (2) to [] node [midway, fill=white, font=\fontsize{8}{0}\selectfont] {$1$} (4);
    
    \draw (2) to [] node [midway, fill=white, font=\fontsize{8}{0}\selectfont] {$1$} (5);

    \draw (3) to [] node [midway, fill=white, font=\fontsize{8}{0}\selectfont] {$1$} (4);
    
    \draw (3) to [] node [midway, fill=white, font=\fontsize{8}{0}\selectfont] {$1$} (5);
    \end{tikzpicture}

    \caption{Optimal solutions of the $\rqubo$ model for the max 3-cut problem with different penalty coefficients. Left: optimal solution with $c_1 = c_3 = 3 + \epsilon$, $c_4 = c_5 = 2 + \epsilon$, and $c_2 = 4 + \epsilon$ that is optimal for the max $k$-cut problem.  Right: optimal solution after changing $c_2$ from $4 + \epsilon$ to $4 - \epsilon$ that is still an optimal solution for the max $k$-cut problem.}\label{fig:max_k_cut rqubo counter example1.5} 
	\end{figure}
	Let $\epsilon = 10^{-3}$. If we set the penalty coefficients $c_1 = c_3 = d_1^+ - 2d_1^- +\epsilon = d_3^+ - 2d_3^- + \epsilon$, $c_4 = c_5 = d_4^+ - 2d_4^- +\epsilon = d_5^+ - 2d_5^- + \epsilon$ and $c_2 = d_2^+ - 2d_2^- + \epsilon = 4 + \epsilon$, 
    satisfying the lower bound for the penalty coefficients in Theorem~\ref{thm:rqubo},
	then the optimal solution of~\eqref{eq:ex4.5} is $x^*$ with objective  $\bar{q}(x^*) = 5$ (see Figure~\ref{fig:max_k_cut rqubo counter example1.5} (left). 
	If we now change $c_2$ from $4 + \epsilon$ to $4 - \epsilon$, violating the lower bound for the penalty coefficients in Theorem~\ref{thm:rqubo}, then 
	the optimal solution of~\eqref{eq:ex4.5} is still $x^*$ with $\bar{q}(x^*) = 5$ (see Figure~\ref{fig:max_k_cut rqubo counter example1.5} (right)).
	Thus, slightly decreasing the penalty coefficient $c_2$ from $4+\epsilon$ to $4-\epsilon$ does not affect the optimality for the max $k$-cut problem of the optimal solutions obtained by the $\rqubo$ formulation.}
    \end{example}

Although Theorem~\ref{thm:rqubo} does not provide the tightest lower bound for the penalty coefficients in the case when the graph of interest contains negative edge weights, we conjecture that the general tightest lower bound is very close to the one provided in Theorem~\ref{thm:rqubo}. Specifically, we conjecture that the tightest penalty coefficients are obtained by changing $-2d^-_v$ to $-d^-_v$ for all $v \in V$ in Theorem~\ref{thm:rqubo}. Positive evidence for this conjecture stems from unreported numerical experiments and Appendix~\ref{sec:appconj2}.

\begin{conjecture}\label{conjecture: optimal maxKcut rqubo formulation}
Let $G(V,E$) be a graph with edge weights $w_{uv}$ for all $\{u,v\} \in E$. Let
$\hat{x} \in \{0,1\}^{n \times (k-1)}$ be an optimal solution for the $\rqubo$ model~\eqref{eq: maxKcut_rqubo} with $c_v > d_v^+ -  d_v^-$ for every vertex $v \in V$.
Then, $\hat{x}$ is an optimal solution of the max $k$-cut problem.
\end{conjecture}

\section{Computational Experiments}\label{sec:num}

In this section, we illustrate our results on QUBO reformulations of the max $k$-cut problem by numerically investigating the performance of these reformulations when using QAOA to solve them.  Experiments are run with {\tt Qiskit}~\citep{wille2019ibm} on {\tt IBM}'s quantum simulators (via {\tt Qiskit Aer})~\citep{QiskitAer}.  All the code and data developed for these experiments are publicly available on the {\tt GitHub} repository \href{https://github.com/AdrianHarkness/Max_K_Cut}{https://github.com/AdrianHarkness/Max\_K\_Cut}.

\subsection{Experimental Setup}
We use {\tt Qiskit's} implementation of QAOA on IBM's quantum simulators to solve the QUBO and R-QUBO reformulations of the max $k$-cut problem using the penalty coefficients characterized in Sections~\ref{sec:BQO_QUBO} and \ref{sec:R-BQO_QUBO}. Specifically, we collect statistics on these solution approaches by considering 10 instances of the max $3$-cut problem generated using Erd\H{o}s-R\'enyi graphs with edge probability $0.5$.  As our QUBO models require $nk$ qubits while our R-QUBO models require $n(k-1)$ qubits, we restrict our (noisy) density matrix simulations to small graph sizes with $n=5$.  All edges on each graph have a weight of one, except for one randomly selected edge with weight $10$, a decision made in order to magnify the difference between the value of {\em tight} penalty coefficients $c_{\rm tight}$ and {\em naive} penalty coefficients $c_{\rm naive}$. Here, $c_{\rm tight}$ denotes the {\em threshold} value in Theorem~\ref{theorem: optimal maxKcut qubo formulation} (QUBO) and Theorem~\ref{thm:rqubo} (R-QUBO), which these results require to be strictly exceeded, so that $c_{\rm tight}$ is the limiting tight value; and $(c_{\rm naive})_v = \max\{n/k, km\}$ for all $v \in V$; that is, the size-based coefficient in~\eqref{eq:cnaive}, used here as a deliberately naive baseline even though it was originally stated for unweighted graphs.  QAOA is run on each graph with depth $p=1$ (one layer of cost and mixer Hamiltonians).

To collect the desired statistics, we proceed as follows. For each graph both the QUBO and the R-QUBO model are constructed with penalty coefficients given by
\begin{equation}
\label{eq:interpol}
c(t) =(1-t)\, c_{\rm tight} + t\, c_{\rm naive},
\end{equation}
for $t \in [0,1]$. That is, with $t=0$ the models use the tight-threshold values and with $t=1$ the models have naive penalty coefficients.  We then optimize the parameters of each QAOA model using the COBYLA algorithm using a maximum of 1000 function evaluations.  We use the Conditional Value-at-Risk (CVaR) objective function proposed in \cite{barkoutsos2020improving} as the cost function for this QAOA parameter optimization loop, as it has been shown to improve the performance of QAOA.  Using CVaR, parameterized by an aggregation parameter $\alpha \in [0,1]$, allows one to optimize over the expected value of the tail of a distribution rather than over the expected value of the entire distribution.  In the context of QAOA, the limit $\alpha \to 0^+$ corresponds to optimizing over the best shot (i.e., the cost of the best sample), while using $\alpha=1$ corresponds to the standard expected value over all shots.  An intermediate value of $\alpha$ balances the emphasis placed on better samples against the smoothing of the objective landscape.  For our experiments, we use $\alpha=0.25$.  Following parameter optimization, the QAOA distribution of each model is sampled $10,000$ times to provide final solution bitstrings to evaluate. All approximation-ratio and feasibility-probability distributions reported in Figure~\ref{fig:results} are drawn from a QAOA quantum state with \emph{fixed} (optimized) parameters; that is, the classical optimization loop is run once per instance, and the shots are subsequently collected from the resulting state without any further parameter updates.

The sample solution provided by each shot is only kept if it is a feasible solution of the problem instance, so as to only analyze the quality of feasible sample solutions. We then track the solution quality on a given problem instance according to the expected value of the approximation ratio achieved by each feasible sample solution.  For conciseness, we refer to this simply as the approximation ratio, noting that it is by construction {\em conditional on feasibility}.  Also, we track the percentage of the total shots that produce a feasible solution to estimate the probability of QAOA returning a feasible sample. We also record, for every shot, the {\em Hamming distance to feasibility}: the minimum number of bit flips required to reach a feasible solution.

\subsection{Model Comparison}
Although it is beyond the scope of the article to numerically benchmark the different approaches available to solve the max $k$-cut problem using quantum devices, it is worthwhile to compare the performance of our QUBO and R-QUBO penalty models against other leading methodologies for enforcing max $k$-cut constraints when using QAOA. Next, we briefly describe some of the methodologies compared here.

\subsubsection{XY-mixer} 
\begin{figure}[htbp]
    \centering
    \begin{minipage}[b]{0.3\textwidth}
        \centering
        \includegraphics[width=\textwidth]{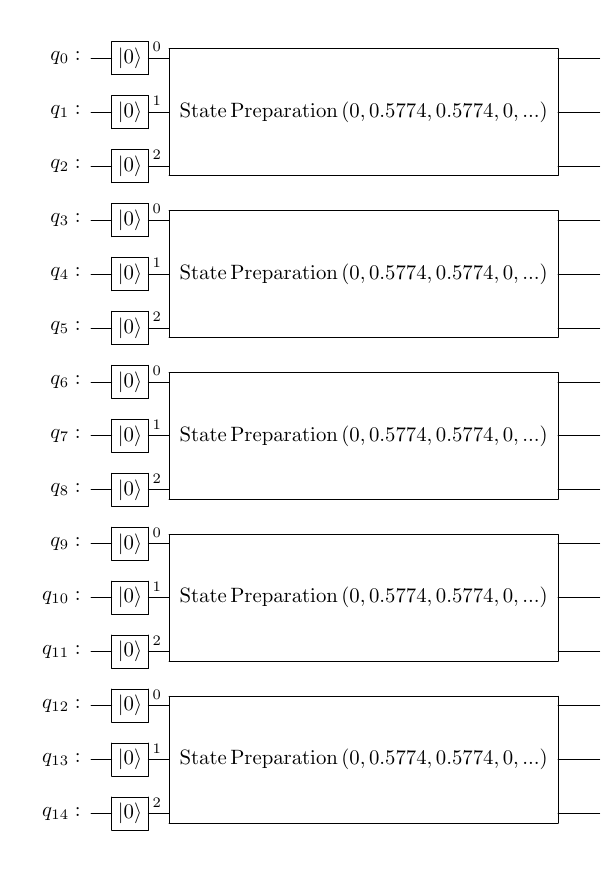}
    \end{minipage}
    \hfill
    \begin{minipage}[b]{0.69\textwidth}
        \centering
        \includegraphics[width=\textwidth]{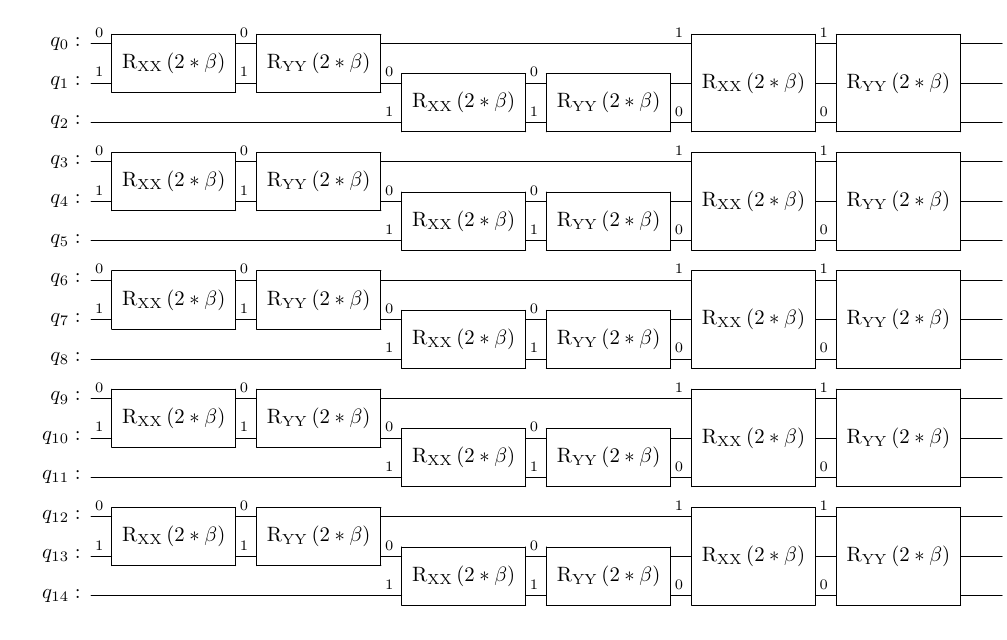}
    \end{minipage}
    \caption{High level view of the quantum circuits used in the XY-mixer formulation for solving the max $k$-cut binary quadratic optimization problem. Left: $D_1^3$ Dicke state preparation. Right: One layer of the XY-mixer.} 
    \label{fig:mixer-circuits}
\end{figure}

As a widely proposed alternative to penalty methods for enforcing Hamming-weight constraints~\citep[see, e.g.,][]{Hadfield2019_from,wang2020xy}, the XY-mixer shifts circuit complexity from the cost Hamiltonian to the mixer Hamiltonian. The mixer is defined as 
\begin{equation}
H_{\text{mixer}} \;=\; \frac{1}{2} \sum_{i,j} \Bigl(X_iX_j \;+\; Y_iY_j\Bigr),
\end{equation}
 where $X_k$ and $Y_k$ denote the Pauli matrices $X$ and $Y$ acting on the $k^{th}$ qubit.  Subspaces spanned by quantum states of equal Hamming weight are invariant under the action of the XY-mixer. Thus, provided an initialization into the desired Hamming-weight subspace defined by problem constraints, the XY-mixer maintains feasibility throughout the QAOA evolution in an idealized noiseless scenario.  Operationally, this means that one can: define a binary quadratic optimization problem with a Hamming-weight equality constraint (such as the one in the BQO formulation~\eqref{eq: maxKcut_qp} using a one-hot constraint); prepare a feasible initial state for QAOA, directly encode the objective into the QAOA cost-Hamiltonian; and use the XY-mixer to enforce the equality constraint.  Often, XY-mixers are paired with initial states known as Dicke states $D_w^n$ that uniformly superpose all $n$-qubit computational basis states with Hamming-weight $w$ (see \cite{Bartschi_2022} for deterministic preparation circuits).  This is analogous to the standard practice of initializing over a uniform superposition over all computational basis states when solving an unconstrained problem.  To solve the max 3-cut problem with an XY-mixer while enforcing that each node only belongs to one partition, we can use disjoint applications of Dicke state preparation circuits for $D_1^3$ on each set of 3 qubits that represent a given node on the graph, which each prepare the state 
\[
\ket{\Psi} = \frac{1}{\sqrt{3}}(\ket{001}+\ket{010}+\ket{100})
\]
across the 3 qubits that they act upon, as well as disjoint XY-mixer circuits on each of those 3-qubit sets that permute Hamming-weight-1 states within the set.  Circuits for these operations are illustrated in Figure \ref{fig:mixer-circuits}.

\subsubsection{Dicke QUBO}
While the standard QUBO and R-QUBO models were initialized in the uniform superposition over all computational basis states, the XY-mixer was initialized in the uniform superposition over feasible states.  This makes it hard to directly compare the two models.  Therefore, we also test a Dicke QUBO model in which: Dicke state preparation is used to initialize over all feasible states; the standard X mixer is used; and the constraints are enforced with a penalized cost Hamiltonian.

\subsubsection{Dicke R-QUBO}
The feasible set of the R-QUBO formulation is not a fixed Hamming-weight subspace: each vertex's register is feasible if and only if its Hamming weight is 0 (partition set $k$) or 1. Dicke states therefore do not directly apply. However, under the variable reduction~\eqref{eq:transform}, the Dicke state $D^k_1$ maps to the $(k-1)$-qubit state
\[
\ket{\Psi} = \tfrac{1}{\sqrt{k}}\Big(\ket{0\cdots 0}
  + \sum_{j \in P \setminus \{k\}} \ket{e_j}\Big)
\]
per vertex, where $\ket{e_j}$ denotes the computational basis state with a single one in position $j$. Since the eliminated variable is a deterministic function of the remaining ones, $\ket{\Psi}$ is the exact uniform superposition over the $k$ feasible states of a single vertex block, and the product state $\ket{\Psi}^{\otimes n}$ is the exact uniform superposition over the full feasible set of the R-QUBO formulation. We therefore refer to the model that combines the $\ket{\Psi}$ initialization, the penalized R-QUBO cost Hamiltonian, and the standard X-mixer as the Dicke R-QUBO. The state $\ket{\Psi}$ is prepared deterministically by a linear cascade: one $R_y$ rotation followed by $k-2$ controlled-$R_y$/CNOT pairs per vertex. General-purpose circuits for preparing Dicke states and, more broadly, symmetric states of Hamming weight at most $k$, are given in \cite{Bartschi_2022}. Note that an XY-mixer is not applicable to this encoding: it preserves each fixed-Hamming-weight subspace and therefore never transfers amplitude to or from the all-zeros state encoding partition set $k$, so the penalty terms remain necessary.

\subsubsection{Penalty+Mixer QUBO}
The last model we test combines all aforementioned elements: we use Dicke state initialization, a cost Hamiltonian with penalty terms, and an XY-mixer Hamiltonian.

\subsection{Results}

\begin{figure}[htbp]
    \centering
    \begin{minipage}{0.99\textwidth}
        \includegraphics[width=\textwidth]{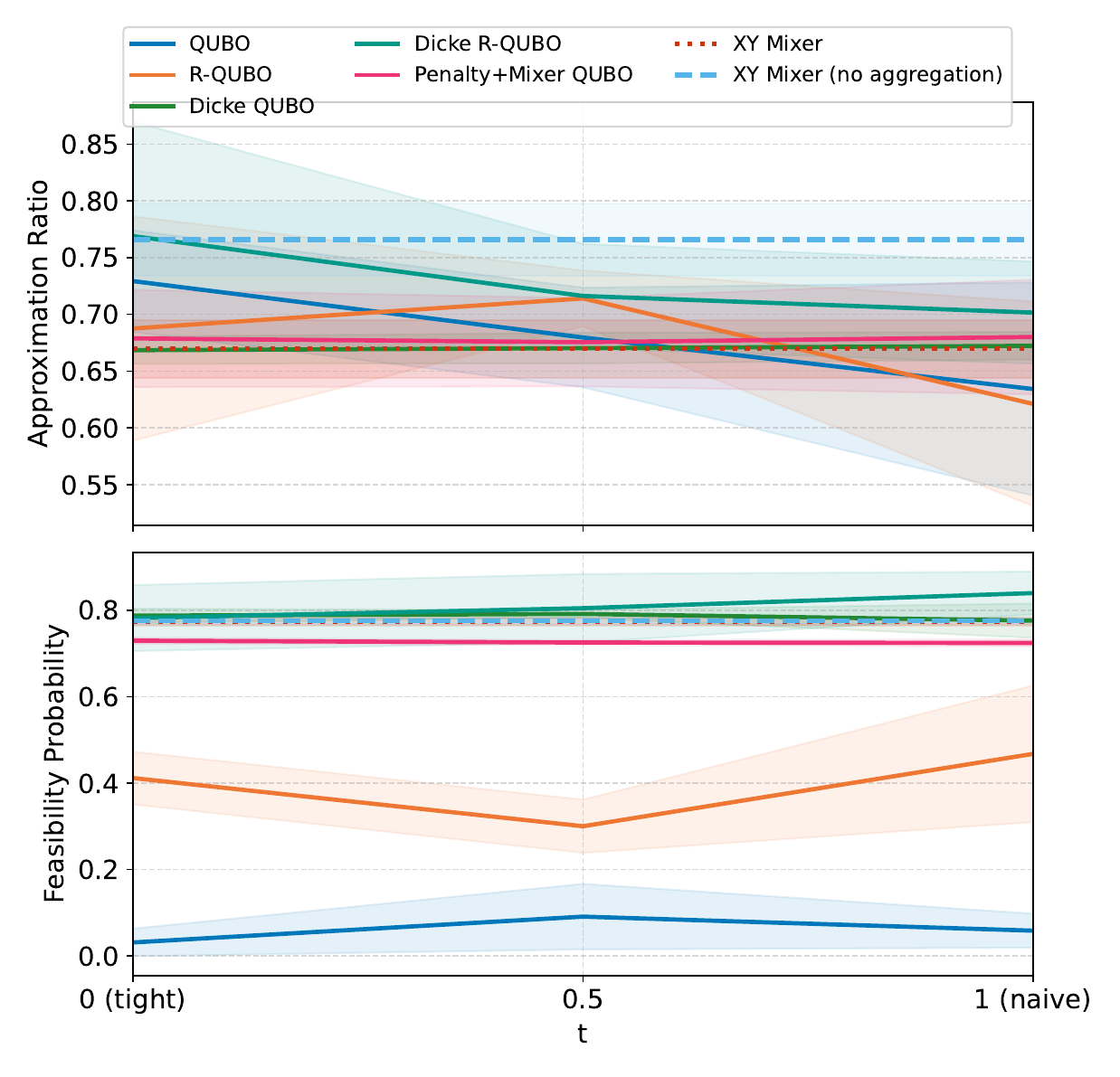}
        \label{fig:noisy}
    \end{minipage}
    \caption{Expected approximation ratio conditional on feasibility, and probability of solution feasibility, for various problem formulations. Expected values for the max $3$-cut problem on Erd\H{o}s-R\'enyi graphs with $n = 5$ nodes and edge probability $0.5$ are obtained after $10,000$ shots of QAOA in a quantum simulator. Dark lines denote expected values over all randomly generated graphs, while shaded regions denote one standard deviation from the mean.} 
    \label{fig:results}
\end{figure}

\begin{figure}[htbp]
    \centering
    \begin{minipage}[b]{0.99\textwidth}
        \centering
        \includegraphics[width=\textwidth]{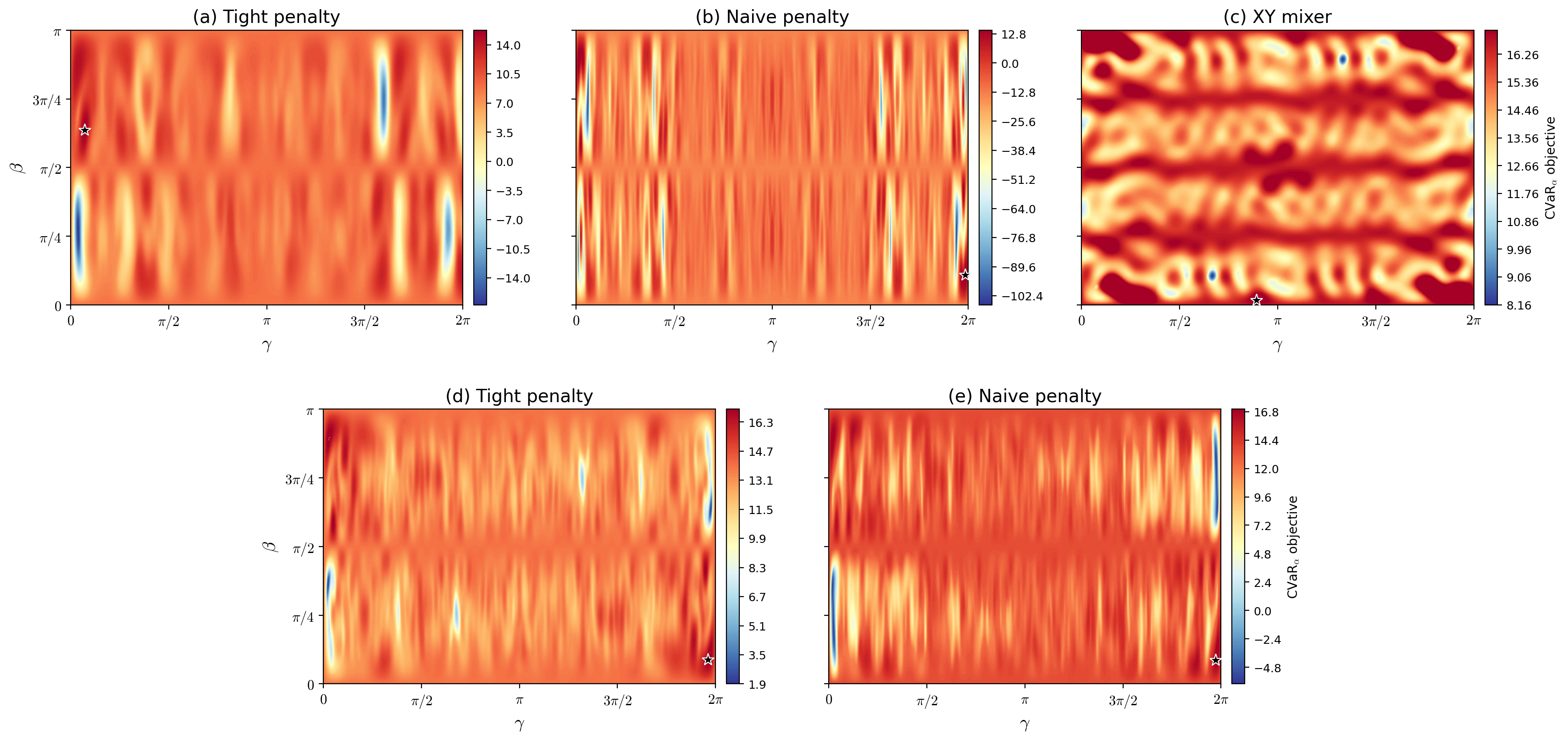}
    \end{minipage}
    \caption{CVaR objective (with $\alpha=0.25$) of the QAOA $p=1$ cost Hamiltonian as a function of $\gamma$ and $\beta$ variational parameters. The models evaluated are (a) QUBO with the tight-threshold penalty, (b) QUBO with a naive penalty, (c) XY-mixer, (d) R-QUBO with the tight-threshold penalty, and  (e) R-QUBO with a naive penalty.  A star marks the optimal point on each plot.}
    \label{fig:cost-landscapes}
\end{figure}

The results obtained from this experimental setting are summarized in Figures~\ref{fig:results} and~\ref{fig:cost-landscapes}, which compare model performance under realistic, noisy simulation of the \texttt{ibm\_boston} quantum computer.

\paragraph{Approximation ratio.} Among all models tested, the highest expected approximation ratios are achieved by the XY-mixer without CVaR aggregation (i.e., not using the CVaR objective during parameter optimization) at ${\approx}0.77$ and the Dicke R-QUBO at the tight-threshold setting at ${\approx}0.77$; the two are essentially tied in mean approximation ratio on this test set (the Dicke R-QUBO attains the higher per-graph value on 6 of 10 instances), with the Dicke R-QUBO doing so on a smaller register and with roughly half the transpiled two-qubit gate count ($98.5 \pm 36.0$ versus $200.8 \pm 45.3$; logical and transpiled counts are reported in Appendix~\ref{app:resources}). Because the XY-mixer formulation does not involve penalty coefficients, it is represented as a single data point in Figure~\ref{fig:results}, plotted as a horizontal line for visual comparison with the penalty-based models. The XY-mixer with CVaR aggregation yields a lower approximation ratio of approximately $0.67$, suggesting that, for this small test set of 10 graphs with $n=5$, CVaR aggregation does not improve solution quality for the XY-mixer as it appears to for the penalty-based formulations. Confirming this observation at larger scales is left for future work.

Among the penalty-based formulations, the QUBO model's conditional approximation ratio decreases monotonically from $0.73$ at $t=0$ to $0.63$ at $t=1$; we caution, however, that this expectation is conditioned on a very small feasible fraction (${\approx}3\%$ of shots at $t=0$), so it reflects strong post-selection rather than reliable feasible sampling. The R-QUBO achieves $0.69$--$0.71$ at the tight and intermediate settings, degrading to $0.62$ with naive penalties. Both trends are consistent with the practical motivation for tight penalties: within each formulation, naive coefficients yield the worst solution quality. The Dicke QUBO remains essentially flat near $0.67$ across all values of $t$, indicating that feasible-subspace initialization mitigates the landscape-flattening effect of naive penalties; the Dicke R-QUBO shows the same flattening benefit at a higher level ($0.77$ at $t=0$, $0.70$ at $t=1$) and outperforms every other penalty-based model at every penalty value. The Penalty+Mixer QUBO, which combines Dicke initialization, penalty terms, and the XY-mixer, is stable near $0.68$, above the QUBO and the XY-mixer with CVaR; this suggests that combining enforcement strategies can outweigh the cost of increased circuit depth, though the Dicke R-QUBO attains both a higher approximation ratio ($0.77$ versus $0.68$) and a higher feasibility probability ($0.78$ versus $0.72$) while using about $39\%$ of its transpiled two-qubit gates.

\paragraph{Feasibility probability.} The feasibility results further distinguish the models. The Dicke R-QUBO achieves the highest feasibility probability of any model tested, rising from approximately $0.78$ at $t=0$ to $0.84$ at $t=1$, with the Dicke QUBO ($0.78$--$0.79$) and both XY-mixer variants (${\approx}0.77$) close behind. Because the XY-mixer formulation does not involve penalty coefficients, these are single data points plotted as horizontal lines. Notably, the observed XY-mixer feasibility is well below the noiseless limit of $1.0$ that would be expected in the idealized setting, since the XY-mixer preserves the Hamming-weight subspace only under exact unitary evolution. Under the noisy simulation of \texttt{ibm\_boston}, gate errors cause leakage out of the feasible subspace, reducing the observed feasibility probability. The XY-mixer with CVaR aggregation performs identically in terms of feasibility probability.

Among the remaining penalty-based models, the R-QUBO attains feasibility between $0.30$ and $0.47$ with no clear monotone dependence on $t$ at this sample size ($0.41$, $0.30$, and $0.47$ at $t=0$, $0.5$, and $1$, with substantial graph-to-graph variance at the naive setting), while the QUBO remains consistently low at $0.03$--$0.09$. Pairing the reduced encoding with feasible-subspace initialization thus roughly doubles the R-QUBO's feasibility while simultaneously improving its conditional solution quality; as with the Dicke QUBO, the initialization is the dominant improving factor. Adding an XY-mixer to the Dicke QUBO again reduces feasibility, to a consistent $0.72$--$0.73$.

Feasibility probability alone does not capture how {\em severely} infeasible samples violate the constraints. Appendix~\ref{app:violation} reports the distribution of the Hamming distance to feasibility: violations are almost exclusively a single bit flip for the Dicke-initialized models (Dicke QUBO and Dicke R-QUBO) and for both XY-mixer variants; they are concentrated within two bit flips for the plain R-QUBO; and they average roughly three for the plain QUBO, essentially independently of $t$. Hence, the penalty magnitude primarily affects how {\em often} the algorithm leaves the feasible space, while the formulation and initialization primarily affect how {\em far} it travels from it.

\paragraph{Effect of penalty magnitude on QAOA energy landscapes.} The penalty magnitude affects the two performance metrics asymmetrically. Its effect on feasibility is weak and non-monotonic at this problem size; as discussed above, the dominant lever on feasibility is the initialization. Its effect on solution quality is more pronounced: every formulation whose approximation ratio varies appreciably with $t$ attains its worst value at the naive setting. This degradation occurs because overly large penalties cause the penalty term to dominate the cost Hamiltonian, which has two compounding effects. First, large penalties compress the relative objective differences among feasible solutions, making it harder for the optimizer to distinguish high-quality from low-quality feasible solutions. Second, in the particular case of QAOA, large penalties introduce high-frequency periodicity into the energy landscape used to optimize the QAOA parameters, creating many local optima in which classical optimizers can become trapped. Both phenomena are much less severe when using tight penalties.

To illustrate this, Figure~\ref{fig:cost-landscapes} plots the CVaR objective of the QAOA $p = 1$ cost Hamiltonian as a function of the QAOA variational parameters $\gamma$ and $\beta$ for a representative graph instance. The naive-penalty QUBO landscape (Figure~\ref{fig:cost-landscapes}(b)) is visibly more periodic in $\gamma$ and contains many more local optima than its tight-penalty counterpart (Figure~\ref{fig:cost-landscapes}(a)). This increased periodicity arises because the large penalty coefficients introduce high-frequency oscillations into the cost Hamiltonian, creating a rugged landscape in which classical optimizers such as COBYLA are more likely to become trapped in suboptimal local optima. The same effect is observed in the R-QUBO landscapes: the tight-penalty landscape (Figure~\ref{fig:cost-landscapes}(d)) exhibits broader, more navigable features, while the naive-penalty landscape (Figure~\ref{fig:cost-landscapes}(e)) displays finer periodic structure and a proliferation of local optima. Additionally, comparing the color bar scales reveals that the naive-penalty landscapes span a much wider range of objective values (e.g., $-102$ to $12.8$ for the naive QUBO versus $-14$ to $14$ for the tight QUBO), reflecting the dominance of the penalty term, which compresses the relative differences among feasible solutions and further hinders the optimizer's ability to distinguish high-quality from low-quality feasible solutions. For this instance, the XY-mixer landscape (Figure~\ref{fig:cost-landscapes}(c)) largely avoids these issues, exhibiting smooth, broad features with a clearly identifiable optimum, consistent with its strong approximation ratio performance.

\subsubsection{Results summary} Among penalty-based approaches, and indeed among all models tested, the Dicke R-QUBO offers the best overall balance: the highest feasible-sample approximation ratio (${\approx}0.77$ at the tight threshold, essentially tied with the XY-mixer without aggregation), the highest feasibility probability ($0.78$--$0.84$), the joint-smallest register, tied with the plain R-QUBO ($n(k-1)$ versus $nk$ qubits), and a transpiled circuit roughly half the size of the Dicke QUBO's on average ($98.5 \pm 36.0$ versus $172.6 \pm 41.9$ two-qubit gates; see Appendix~\ref{app:resources}). Its tight-threshold penalties retain the landscape advantages observed for the R-QUBO in Figure~\ref{fig:cost-landscapes}, while the feasible-subspace initialization removes the feasibility cost of the penalty approach. More broadly, Dicke-type initialization is the most impactful single ingredient among the models tested, and the reduced encoding makes it cheaper: preparing the at-most-one-hot superposition requires 10 logical ($13.0 \pm 3.0$ transpiled) two-qubit gates versus 20 logical ($32.9 \pm 4.2$ transpiled) for one-hot Dicke preparation. Notably, augmenting the Dicke QUBO with an XY-mixer actually reduces feasibility, which is consistent with greater noise-induced leakage offsetting the mixer's theoretical feasibility guarantees; the mixer adds roughly $71$ transpiled two-qubit gates per layer within the mixer segment (plus routing overhead elsewhere in the circuit).

The XY-mixer without CVaR aggregation matches the best approximation ratio (${\approx}0.77$) but requires roughly twice the two-qubit gates of the Dicke R-QUBO and offers no feasibility advantage under noise. We also observe that CVaR aggregation does not uniformly improve performance: for the XY-mixer, it degrades the approximation ratio ($0.67$ versus $0.77$) without improving feasibility, suggesting that practitioners should evaluate whether CVaR benefits their specific formulation rather than applying it by default.

It is important to emphasize that these results are only illustrative in nature and should only serve as motivation to perform future comprehensive numerical experiments (that are beyond the scope of this article) that could better test these conclusions. These illustrative experiments also leave untested the potential performance of penalization methods on adiabatic quantum computing machines, where other related articles have found the methodology to perform very well~\citep[see, e.g.,][]{brown2024copositive,quintero2021_characterization}.

\section{Final Remarks}
\label{sec:fin}
Building on recent literature focused on deriving QUBO reformulations of combinatorial optimization problems that are tailored to leverage the capabilities of quantum computing devices, this article presents a characterization of tight penalty coefficients for QUBO reformulations of the max $k$-cut problem. To illustrate our results, we perform a small sample of numerical experiments in which the QUBO reformulations are used to solve the max $k$-cut problem using the QAOA algorithm. However, as mentioned earlier, our results are relevant to address the solution of the max $k$-cut problem using adiabatic quantum computing and Ising machines which have shown promising numerical results in solving QUBO reformulations of combinatorial optimization problems of appreciable scale~\citep[see, e.g.,][]{brown2024copositive,quintero2021_characterization}. 
Also, our results allow for the use of the quantum Hamiltonian descent (QHD) algorithm~\citep{leng2023quantum}, as the binary constraints in the QUBO reformulations can be relaxed following the approach in~\cite{rosenberg1972breves}.

Furthermore, our results are also relevant when considering the solution of the max $k$-cut problem with classical optimization solvers by following an approach analogous to the one used in~\cite{abello2001finding} and \cite{pardalos2006continuous} for other combinatorial problems. 

We finish by mentioning that in this work, we considered a standard binary encoding of the max $k$-cut problem. However, there are more efficient binary encodings (in terms of the number of binary variables required) of the max $k$-cut, as recently studied in~\cite{fuchs2025encodings}. It will be interesting to investigate (when relevant) the tightest penalty coefficients in these settings. In some cases --- as for the knapsack problem~\citep[][Cor. 2]{quintero2021KP} --- tight penalty coefficients that are valid in one binary encoding are valid for a different binary encoding.

\section*{Acknowledgments} This research project has been carried out thanks to funding by (i) the Defense Advanced Research
Projects Agency (DARPA), ONISQ grant W911NF2010022, titled The Quantum Computing Revolution and Optimization: Challenges and Opportunities, and (ii) the Office of Naval Research under grant number N000142412648. Also, this research used resources of the Oak Ridge Leadership Computing Facility, which is a DOE Office of Science User Facility supported under Contract DE-AC05-00OR22725. This research was supported by PNNL's Quantum Algorithms and Architecture for Domain Science (QuAADS) Laboratory Directed Research and Development (LDRD) Initiative. This material is based upon work supported by the
U.S. Department of Energy, Office of Science, National Quantum Information Science Research Centers, Quantum Science
Center (QSC). The Pacific Northwest National Laboratory
is operated by Battelle for the U.S. Department of Energy
under Contract No. DE-AC05-76RL01830.

\bibliographystyle{unsrtnat}
\bibliography{max_k_cut}

\appendix

\section{Proof of Lemma~\ref{lem:rqubo}}
\label{app:lemma2proof}

\begin{proof}[Proof of Lemma~\ref{lem:rqubo}]

For any $x \in \{0,1\}^{n\times (k-1)}$, and any vertex pair $\{u, v\} \in \binom{V}{2}$, let 
\begin{align*}
t_{uv}(x) : = \sum_{j \in \bar{P}} x_{uj} x_{vj} && \text{and} &&  t_{u}(x) : = \sum_{j \in \bar{P}} x_{uj}.
\end{align*}
Recall the $\rbqo$ formulation of the max $k$-cut problem, and note that if $t_u(x) = 0$, then vertex $u$ is assigned to set $k$ in the partition $P$, and if  $t_u(x) = 1$, then vertex $u$ is assigned to some set $j\in \{1, \dots, k-1\}$ in the partition $P$. Thus, from the $\bqo$ formulation of the max $k$-cut problem, to prove the result, it is enough to show that $t_v(\hat{x})\le 1$ for all vertices $v \in V$.

Also, let
\begin{align}
\bar{q}_1(x): = & - \sum_{\{u,v\} \in E} w_{uv} \left (t_{uv}(x) + \left (1 -t_{u}(x)\right) \left(1 - t_{v}(x) \right) \right)
& \bar{q}_2(x) := & -\sum_{v \in V} c_v \sum_{\{i,j\} \in \binom{\bar{P}}{2}} x_{vi} x_{vj}
\label{eq:q1} 
\end{align}
Thus, it follows that $\bar{q}(x) = \sum_{\{u,v\} \in E} w_{uv} + \bar{q}_1(x) + \bar{q}_2(x)$. 

\looseness -1
By contradiction, assume that $\hat{x}$ is not feasible for the max $k$-cut problem. Then, it follows that there is $u \in V$ such that $t_u(\hat{x}) > 1$. In particular, let $s \in \argmax_{u \in V}\{t_u(\hat{x}): t_u(\hat{x}) > 1\}$.
Now, let $\bar{x}(s) \in \{0,1\}^{n \times (k-1)}$ be defined as follows. For any $v \in V$, $v \neq s$, let $\bar{x}_{vj}(s) = \hat{x}_{vj}$ for all $j \in \bar{P}$.
Also, choose $j' \in \bar{P}$, such that $\hat{x}_{sj'} = 1$ and let $\bar{x}_{sj}(s) = 0$ for all $j \in \bar{P} \setminus \{j'\}$, and $\bar{x}_{sj'}(s) = 1$. Clearly, $\bar{x}(s)$ is feasible for $\rqubo$. We claim that $\bar{q}(\bar{x}(s)) > \bar{q}(\hat{x})$, contradicting the optimality of $\hat{x}$, and therefore the statement follows.

To prove the claim, which is equivalent to showing that $(\bar{q}_1(\bar{x}(s)) - \bar{q}_1(\hat{x})) +(\bar{q}_2(\bar{x}(s)) - \bar{q}_2(\hat{x})) > 0$, we proceed as follows. First, for the purpose of clarity, for any $u, v \in V$, let $\hat{t}_{uv} = t_{uv}(\hat{x})$, $\hat{t}_u = t_u(\hat{x})$, and $\hat{t}_v = t_v(\hat{x})$. Similarly, for any $u, v \in V$, let $\bar{t}_{uv} = t_{uv}(\bar{x}(s))$, $\bar{t}_u = t_u(\bar{x}(s))$, and $\bar{t}_v = t_v(\bar{x}(s))$. In particular, $\hat{t}_s >1$ and $\bar{t}_s =1$. Without loss of generality, we assume that for any $v \in N_G(s)$, the edge between $v$ and $s$ in the set $E$ is labeled $\{s,v\}$.

Now we proceed to find an appropriate bound on $\bar{q}_1(\bar{x}(s)) - \bar{q}_1(\hat{x})$. For that purpose, for any $x \in \{0,1\}^{n \times (k-1)}$, and $u,v \in V$, let 
\[
h_{uv}(x) = w_{uv} \left (t_{uv}(x) + \left (1 -t_{u}(x)\right) \left(1 - t_{v}(x) \right) \right).
\]
Again, for brevity, we let $\hat{h}_{uv} = h_{uv}(\hat{x}) = w_{uv}(\hat{t}_{uv} + (1-\hat{t}_u)(1-\hat{t}_v))$, and $\bar{h}_{uv} = h_{uv}(\bar{x}(s)) = w_{uv}(\bar{t}_{uv} + (1-\bar{t}_u)(1-\bar{t}_v))$. From~\eqref{eq:q1}, it follows that $\bar{q}_1(\bar{x}(s)) - \bar{q}_1(\hat{x}) = \sum_{\{s,v\} \in E} (\hat{h}_{sv} - \bar{h}_{sv})$. Thus, we next proceed by setting bounds on $\hat{h}_{sv} - \bar{h}_{sv}$ for all  $\{s,v\} \in E$ in a case-by-case basis. In all these cases, we will use the fact that since $\bar{t}_s = 1$, then for any $\{s,v\} \in E$, $\bar{h}_{sv} = w_{sv} \bar{t}_{sv}$; and in particular, if $w_{sv} > 0$ (resp. $w_{sv} < 0$), then $\bar{h}_{sv} \ge 0$ (resp. $\bar{h}_{sv} \le 0$).

If $\{s,v\} \in E^+$ (i.e., $w_{sv} > 0$) for any $v \in V$, consider three cases:
\begin{enumerate}[label = (\roman*)]
\item $\hat{t}_v >1$: Since $\hat{t}_v >1$, $\hat{t}_s >1$, then $\hat{h}_{sv} \ge w_{sv}(0 + (1-2)(1-2)) = w_{sv}$. Also, since $\bar{t}_s = 1$ implies $\bar{t}_{sv} \le 1$, then $\bar{h}_{sv} \le w_{sv}$. Thus, $\hat{h}_{sv} - \bar{h}_{sv} \ge w_{sv} - w_{sv} = 0 \ge w_{sv}(1 - \hat{t}_s)$.
\item  $\hat{t}_v =1$: Since $\hat{t}_v =1$, $\hat{t}_s >1$, then $\hat{h}_{sv} \ge w_{sv}(\hat{t}_{sv} + (2-1)(1-1)) = w_{sv}\hat{t}_{sv} $. Also, from the construction of~$\bar{x}(s)$ (i.e., the partition assignments of $v$ are not changed), it follows that $\bar{t}_{sv} \le \hat{t}_{sv}$. Therefore, $\bar{h}_{sv} = w_{sv}\bar{t}_{sv} \le w_{sv}\hat{t}_{sv}$.
Thus, $\hat{h}_{sv} - \bar{h}_{sv} \ge w_{sv}\hat{t}_{sv}  - w_{sv}\hat{t}_{sv}  = 0\ge w_{sv}(1 - \hat{t}_s)$.
\item  $\hat{t}_v =0$: Since $\hat{t}_v =0$, then $\hat{t}_{sv} = 0$. This, together with $\hat{t}_s >1$, implies that $\hat{h}_{sv} \ge w_{sv}(0 + (1-\hat{t}_s)(1-0)) = w_{sv}(1 - \hat{t}_{s})$. 
From the construction of $\bar{x}(s)$, it follows that $\bar{t}_v = 0$, which implies that $\bar{t}_{sv}  = 0$. Therefore, $\bar{h}_{sv}  = w_{sv}\bar{t}_{sv} = 0$.
Thus, $\hat{h}_{sv} - \bar{h}_{sv} = w_{sv}(1 - \hat{t}_{s})$.
\end{enumerate}
From these cases, we can conclude that
\begin{align}
\begin{split}
\label{eq:plusedges}
\sum_{\{s,v\} \in E^+} (\hat{h}_{sv} - \bar{h}_{sv}) \ge -\sum_{\{s,v\} \in E^+} w_{sv}(\hat{t}_{s}-1) &\ge -\sum_{\{s,v\} \in E^+} w_{sv}\frac{(\hat{t}_{s}-1)\hat{t}_{s}}{2}  \\
& = -\sum_{v \in N^+_G(s)} w_{sv} \binom{\hat{t}_{s}} {2} = -d_s^+\binom{\hat{t}_{s}}{2}, 
\end{split}
\end{align}
where in the last inequality, we use again that $\hat{t}_s >1$ (i.e., $\hat{t}_s \ge 2$).

Now, if $\{s,v\} \in E^-$ (i.e., $w_{sv} < 0$) for any $v \in V$, consider four cases, where for any $v \in V$, we let $\bar{P}_v := \{ j \in \bar{P}: \hat{x}_{vj} = 1\}$:
\begin{enumerate}[label = (\roman*), wide, labelindent=0pt]
\item $\hat{t}_v >1$ and $\bar{P}_s \subseteq \bar{P}_v$:  From $\bar{P}_s \subseteq \bar{P}_v$, 
and the definition of node $s$ (c.f., beginning of the proof), 
it follows that $\bar{P}_s \equiv \bar{P}_v$. In particular,  
$\hat{t}_{sv} = \hat{t}_s = \hat{t}_v$; which implies that $\hat{h}_{sv} = w_{sv}(\hat{t}_s +1 - \hat{t}_s - \hat{t}_s +{\hat{t}_s}^2) = w_{sv}(\hat{t}_s(\hat{t}_s - 1) + 1)$. 
Also, from $\bar{P}_s \equiv \bar{P}_v$ and the construction of $\bar{x}(s)$, it follows that $\bar{t}_{sv} = 1$, which implies that $\bar{h}_{sv} = w_{sv}$.
Thus, $\hat{h}_{sv} - \bar{h}_{sv} = w_{sv}\hat{t}_s(\hat{t}_s - 1)$.

\item $\hat{t}_v >1$ and $\bar{P}_s \not \subseteq \bar{P}_v$:  From $\bar{P}_s \not \subseteq \bar{P}_v$, it follows that $\hat{t}_{sv} \le \hat{t}_s - 1$. Further, from the definition of node $s$, it follows that $\hat{t}_s \ge \hat{t}_v$. Therefore, $\hat{h}_{sv} = |w_{sv}|(-(\hat{t}_{sv} - \hat{t}_s) - (1-\hat{t}_v + \hat{t}_s\hat{t}_v)) \ge |w_{sv}|(-(-1) - (1-\hat{t}_v+ \hat{t}_s\hat{t}_v)) = w_{sv}\hat{t}_v(\hat{t}_s - 1)$. Thus, $\hat{h}_{sv} - \bar{h}_{sv} \ge w_{sv}\hat{t}_v(\hat{t}_s - 1) - w_{sv}\bar{t}_{sv} \ge w_{sv}\hat{t}_s(\hat{t}_s - 1)$.

\item $\hat{t}_v = 1$: From $\hat{t}_v = 1$, it follows that $\hat{t}_{sv} \le 1$; which implies that $\hat{h}_{sv} = w_{sv}(\hat{t}_{sv} + (1-\hat{t}_s)(1-1)) = w_{sv}\hat{t}_{sv} \ge w_{sv} \ge w_{sv}(\hat{t}_s -1)$. Thus, $\hat{h}_{sv} - \bar{h}_{sv} \ge w_{sv}(\hat{t}_s -1)- w_{sv}\bar{t}_{sv} \ge  w_{sv}\hat{t}_s(\hat{t}_s -1)$.
\item $\hat{t}_v = 0$: From $\hat{t}_v = 0$, it follows that $\hat{t}_{sv} = 0$; which implies that $\hat{h}_{sv} = w_{sv}(0 + (1-\hat{t}_s)(1-0)) = w_{sv}(1- \hat{t}_{s}) \ge 0 \ge w_{sv}(\hat{t}_{s}-1) $. Thus, $\hat{h}_{sv} - \bar{h}_{sv} \ge  w_{sv}\hat{t}_s(\hat{t}_s -1) - w_{sv}\bar{t}_{sv} \ge w_{sv}\hat{t}_s(\hat{t}_s -1)$.
\end{enumerate}
From these cases, we can conclude that
\begin{align}
\begin{split}
\label{eq:minusedges}
\sum_{\{s,v\} \in E^-} (\hat{h}_{sv} - \bar{h}_{sv}) \ge \sum_{\{s,v\} \in E^-} w_{sv}\hat{t}_{s}(\hat{t}_{s}-1) 
&= 2\sum_{\{s,v\} \in E^-} w_{sv}\frac{(\hat{t}_{s}-1)\hat{t}_{s}}{2}\\
&= 2\sum_{v \in N^-_G(s)} w_{sv} \binom{\hat{t}_{s}}{2} = 2d_s^-\binom{\hat{t}_{s}}{2}.
\end{split}
\end{align}
Now, we proceed to find an appropriate bound on $\bar{q}_2(\bar{x}(s)) - \bar{q}_2(\hat{x})$. From the construction of $\bar{x}(s)$, and in particular, from the fact that $\bar{x}_{si}(s)\bar{x}_{sj}(s) = 0$ for all $i\neq j$, $i, j \in \bar{P}$, it follows that
\begin{equation}
\label{eq:q2-new}
\bar{q}_2(\bar{x}(s)) - \bar{q}_2(\hat{x}) = -c_s \sum_{\{i,j\} \in \binom{\bar{P}}{2}} (\bar{x}_{si}(s)\bar{x}_{sj}(s) - \hat{x}_{si}\hat{x}_{sj}) = c_s \binom{\hat{t}_s}{2} > (d_s^+ - 2 d_s^-)\binom{\hat{t}_s}{2}.
\end{equation}
Combining equations~\eqref{eq:plusedges}, \eqref{eq:minusedges}, and~\eqref{eq:q2-new}, we get that
\begin{equation}
\label{eq:allq}
\bar{q}(\bar{x}(s)) - \bar{q}(\hat{x})= (\bar{q}_1(\bar{x}(s)) - \bar{q}_1(\hat{x})) +(\bar{q}_2(\bar{x}(s)) - \bar{q}_2(\hat{x})) > -d_s^+\binom{\hat{t}_{s}}{2} + 2d_s^-\binom{\hat{t}_{s}}{2} + (d_s^+ - 2d_s^-)\binom{\hat{t}_s}{2} = 0;
\end{equation}
which proves the claim.

\end{proof}

\section{Example for Conjecture~\ref{conj:qubo}}
\label{sec:appconj}

The following example shows that if Conjecture~\ref{conj:qubo} holds, then it likely provides the tightest penalty coefficients for general graphs (i.e., with potential negative edge weights). That is, it provides an instance for the max $k$-cut problem with negative edge weights for which violating the lower bound on the penalty coefficients provided in Conjecture~\ref{conj:qubo}, by any small value, results in the optimal solution of the corresponding $\qubo$ reformulation of the max $k$-cut problem not being a feasible solution for the max $k$-cut problem.

\begin{example}\label{example: max_k_cut qubo counter example pos2}
{\em     Figure~\ref{fig:max_k_cut qubo counter example} illustrates an instance of the max 3-cut problem with the optimal objective value of 7. 
	The non-zero entries of an optimal solution $x^* \in \{0,1\}^{5 \times 3}$ for the max 3-cut problem are $x_{11}^* = x_{31}^* = x_{41}^* = 1$, $x_{22}^* = 1$, and $x^*_{53} = 1$. See Figure~\ref{fig:max_k_cut qubo counter example} (left) for an illustration.
	The corresponding $\qubo$ formulation for this problem is as follows.
	\begin{align}
    \label{eq:quboex3}
		\max_{x \in \{0,1\}^{5 \times 3}} q(x) := 7 -  \sum_{\{u,v\} \in E} w_{uv}\left(x_{u1}x_{v1} + x_{u2}x_{v2} + x_{u3}x_{v3}\right) - \sum_{v \in V} c_v \left(x_{v1}+x_{v2}+x_{v3} - 1 \right)^2. 
	\end{align}

	\begin{figure}[ht!]
		\centering
		\includegraphics[scale = 1]{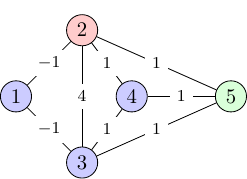}
		 \hspace{8mm}
		\includegraphics[scale = 1]{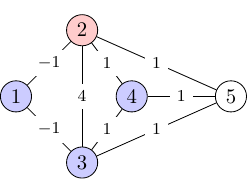}
		 \hspace{8mm}
		\includegraphics[scale = 1]{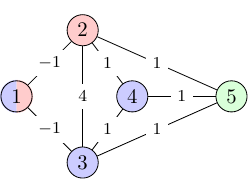}

		\caption{Optimal solutions of the $\qubo$ formulation for the max 3-cut problem with different penalty coefficients. Left: optimal solution with $c_1 = c_5 = 1 + \epsilon$ that is optimal for the max $k$-cut problem. Center:  optimal solution  with $c_1 = 1 + \epsilon$ and $c_5 = 1 -\epsilon$ that is infeasible for the max $k$-cut problem. Right: optimal solution with $c_1 = 1 - \epsilon$ and $c_5 = 1 + \epsilon$ that is infeasible for the max $k$-cut problem.
		}\label{fig:max_k_cut qubo counter example}
	\end{figure}

	Let $\epsilon = 10^{-3}$. If we set the penalty coefficients to $c_v = \max\big\{ \frac{d^+_v}{k},  -\frac{d_v^-}{2
    } \big\} + \epsilon$ for every vertex $v \in V$; that is, we set $c_1 = -\frac{-2}{2} + \epsilon = 1 + \epsilon$, $c_2 = c_3 = \max\big\{\frac{6}{3}, \frac{1}{2}\big\} + \epsilon =2 + \epsilon$, and $c_4 = c_5 = \frac{3}{3} + \epsilon = 1 + \epsilon$, satisfying the lower bound for the penalty coefficients in Conjecture~\ref{conj:qubo},
then the optimal solution of~\eqref{eq:quboex3} is $x^*$ with objective $q(x^*) = 7$ (see Figure~\ref{fig:max_k_cut qubo counter example} (left)).
	If we now change $c_5$ from $1 + \epsilon$ to $1- \epsilon $,
    violating the lower bound for the penalty coefficients in Conjecture~\ref{conj:qubo},
	then the non-zero entries of an optimal solution $\hat{x} \in \{0,1\}^{5 \times 3}$ for the $\qubo$ formulation~\eqref{eq:quboex3} are $\hat{x}_{11} = \hat{x}_{31} = \hat{x}_{41} = 1$ and $\hat{x}_{22} = 1$ with $q(\hat{x})  = 7+\epsilon$.
	In fact, $\hat{x}_{51} = \hat{x}_{52} = \hat{x}_{53} = 0$ (see Figure~\ref{fig:max_k_cut qubo counter example} (center)). 
	Thus, $\hat{x}$ is an infeasible solution for the max 3-cut problem and $q(\hat{x}) > q(x^*)$. 
	Similarly, if we change $c_1$ from $1 + \epsilon$ to $1- \epsilon $, then the non-zero entries of an optimal solution $\tilde{x}$ for the $\qubo$ formulation~\eqref{eq:quboex3} are $\tilde{x}_{11} = \tilde{x}_{12} = 1$ with $q(\tilde{x})  = 7+\epsilon$ (see  Figure~\ref{fig:max_k_cut qubo counter example} (right)). Thus, $\tilde{x}$ is not feasible for the max $k$-cut problem and $q(\tilde{x}) > q(x^*) $. }
\end{example}

\section{Example for Conjecture~\ref{conjecture: optimal maxKcut rqubo formulation}}
\label{sec:appconj2}

The following example shows that if Conjecture~\ref{conjecture: optimal maxKcut rqubo formulation} holds, then it likely provides the tightest penalty coefficients for general graphs (i.e., with potential negative edge weights). That is, it provides an instance for the max $k$-cut problem with negative edge weights for which violating the lower bound on the penalty coefficients provided in Conjecture~\ref{conjecture: optimal maxKcut rqubo formulation}, by any small value, results in the optimal solution of the corresponding $\rqubo$ reformulation of the max $k$-cut problem not being a feasible solution for the max $k$-cut problem.
  
\begin{example}\label{example:rqubo counter example2} 
{\em Figure~\ref{fig:max_k_cut rqubo counter example2} illustrates an instance of the max 3-cut problem with optimal objective value of 6. 
The non-zero entries of an optimal solution $x^* \in \{0,1\}^{5 \times 2}$ of the max 3-cut problem are $x_{21}^* = x_{31}^* = 1$,  and $ x_{52}^* = 1$. See Figure~\ref{fig:max_k_cut rqubo counter example2} (left) for an illustration.
The corresponding $\rqubo$ formulation for this problem is as follows
	\begin{equation}
    \label{eq:ex6}
	    \bar{q}(x) =5 - \sum_{\{u,v\} \in E} w_{uv} \Big[ \sum_{j \in \{1,2\} } x_{uj}x_{vj} + \Big(1 -\sum_{j \in \{1,2\}} x_{uj}\Big)  \Big(1 -\sum_{j \in \{1,2\}} x_{vj}\Big) \Big] - \sum_{v \in V} c_v x_{v1} x_{v2}.
	\end{equation}

    \begin{figure}[ht!]
		\centering
		\includegraphics[scale = 1]{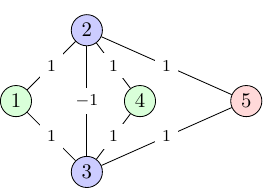}
		 \hspace{8mm}
		\includegraphics[scale = 1]{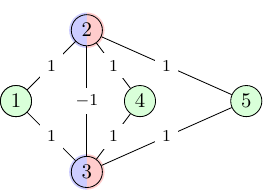}

    \caption{Optimal solutions of the $\rqubo$ model for the max 3-cut problem with different penalty coefficients: (left) an optimal solution $x^*$ of the max $3$-cut problem with $c_1 = c_4 = c_5 = 2 + \epsilon$ and $c_2 = c_3 = 4 + \epsilon$, and (right) an infeasible solution $\hat{x}$ of the $\rqubo$ model after changing $c_2$ from $4 + \epsilon$ to $4 - \epsilon$.}\label{fig:max_k_cut rqubo counter example2} 
	\end{figure}

	Let $\epsilon = 10^{-3}$. If we set the penalty coefficients $c_2 = c_3 = d_2^+ - d_2^- + \epsilon= d_3^+ - d_3^- + \epsilon = 4 + \epsilon$ and $c_v = d_v^+ - d_v^- + \epsilon = 2 + \epsilon$ for every vertex $v \in \{1,4,5\}$, satisfying the lower bound for the penalty coefficients in Conjecture~\ref{conjecture: optimal maxKcut rqubo formulation}, 
	then the optimal solution of~\eqref{eq:ex6} is $x^*$ with  $\bar{q}(x^*) = 6$ (see Figure~\ref{fig:max_k_cut rqubo counter example2}  (left)). If we 
	now change $c_2$ from $4 + \epsilon$ to $4 - \epsilon$,
    violating the lower bound for the penalty coefficients in Conjecture~\ref{conjecture: optimal maxKcut rqubo formulation}, then the non-zero entries of an optimal solution $\hat{x} \in \{0,1\}^{5 \times 2}$ for the $\rqubo$ formulation~\eqref{eq:ex6} are $\hat{x}_{21} = \hat{x}_{22} = 1$, $\hat{x}_{31} = \hat{x}_{32} = 1$ with $\bar{q}(\hat{x}) = 6 + \epsilon$. In fact, 
    $\hat{x}_{vj} = 0$ for $v\in \{1, 4, 5\}$ and $j \in \{1, 2\}$ (see Figure~\ref{fig:max_k_cut rqubo counter example2}  (right)). Thus, $\hat{x}$ is an infeasible solution for the max 3-cut problem and $\bar{q}(\hat{x})>\bar{q}(x^*)$.}
\end{example}

\section{Magnitude of constraint violations}
\label{app:violation}

All the experiments reported in this appendix and in Appendix~\ref{app:resources} were obtained with {\tt Qiskit}~1.3.2, {\tt Qiskit Aer}~0.15.1, {\tt Qiskit Algorithms}~0.3.1, and {\tt Qiskit Optimization}~0.6.1, with the noise model constructed from a snapshot of the {\tt ibm\_boston} backend.

\begin{figure}[htbp]
    \centering
    \includegraphics[width=0.99\textwidth]{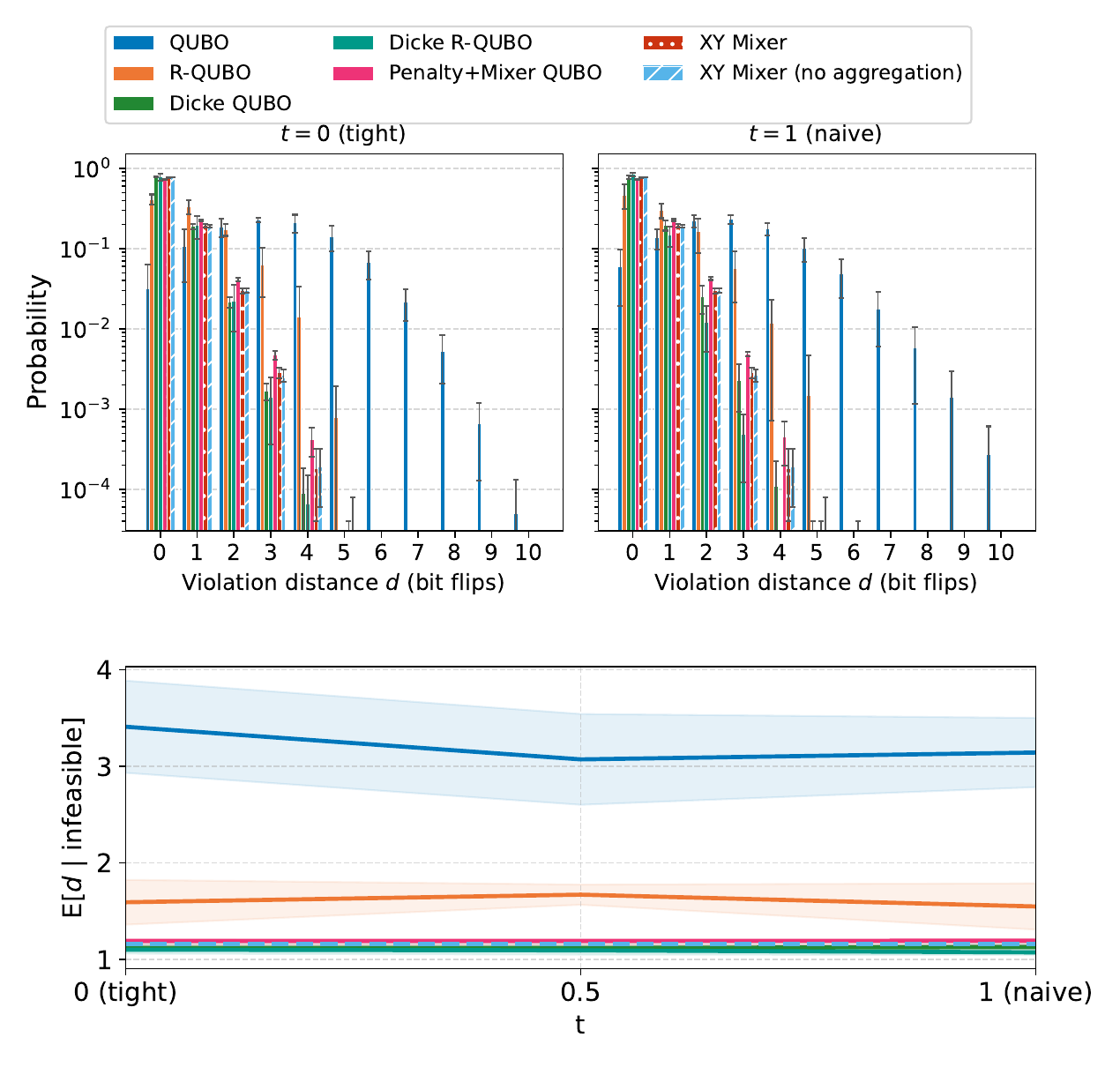}
    \caption{Distribution of the Hamming distance to feasibility (minimum bit flips to the nearest feasible solution) for sampled bitstrings, under noisy simulation of {\tt ibm\_boston}. Top: probability mass at each violation distance $d$ for tight ($t=0$, left) and naive ($t=1$, right) penalty coefficients; bars and error bars denote the mean and one standard deviation over the 10 random graphs, and the $d=0$ mass equals the feasibility probability of Figure~\ref{fig:results}. The XY-mixer models do not depend on $t$ and are identical in both panels. Bottom: expected violation distance conditional on infeasibility as a function of $t$. $\rqubo$ distances are structurally bounded by $n(k-2)=5$ versus $n(k-1)=10$ for the one-hot encodings.}
    \label{fig:violation_distance}
\end{figure}

The feasibility probabilities reported in Figure~\ref{fig:results} indicate how \emph{often} each formulation returns an infeasible sample, but not how \emph{severely} such samples violate the problem constraints. To quantify this, we compute for each sampled bitstring its {\em Hamming distance to feasibility}: the minimum number of bit flips required to reach a feasible solution. Because the constraints act on each vertex independently, this distance decomposes over vertices; a vertex whose one-hot block has row sum $s$ contributes $|s-1|$ flips in the $\qubo$ encoding, and $\max(s-1,0)$ flips in the $\rqubo$ encoding. Note that the per-vertex violation is therefore bounded by $k-1$ in the $\qubo$ encoding but by $k-2$ in the $\rqubo$ encoding, so $\rqubo$ violation distances are structurally bounded below those of the one-hot models. Figure~\ref{fig:violation_distance} reports the distribution of this violation distance at tight ($t=0$) and naive ($t=1$) penalties, together with the expected distance conditional on infeasibility, $\mathbb{E}[d \mid d>0]$, as a function of $t$.

The models separate into three regimes. For the Dicke $\qubo$, Dicke $\rqubo$, Penalty+Mixer $\qubo$, and both XY-mixer variants, infeasible samples lie almost entirely one bit flip from feasibility ($83$--$91\%$ of the infeasible mass at $d=1$, with $\mathbb{E}[d \mid d>0] \approx 1.1$--$1.2$ and at most $2\%$ of infeasible mass at $d\ge 3$). This profile is consistent with isolated noise-induced bit flips on otherwise feasible states rather than with the algorithm populating the infeasible subspace. Notably, the Dicke $\qubo$, which uses feasible-subspace initialization with a standard $X$-mixer, matches the XY-mixer profile, suggesting that initialization, rather than the constraint-preserving mixer, may be the primary driver of violation depth under noise. The Dicke $\rqubo$ provides a direct test of this attribution in a setting where the standard XY-mixer is not applicable: feasible-subspace initialization alone, on the reduced encoding, reproduces the shallow-violation profile ($\mathbb{E}[d \mid d>0] \approx 1.1$, with $91\%$ of infeasible mass at $d=1$), whereas the identically penalized $\rqubo$ without it averages ${\approx}1.6$. The $\rqubo$ occupies an intermediate regime ($\mathbb{E}[d \mid d>0] \approx 1.6$, with $88\%$ of infeasible mass at $d\le 2$), so its infeasible samples remain inexpensive to repair. In contrast, the $\qubo$ model's infeasible samples average roughly three vertex-level violations ($\mathbb{E}[d \mid d>0] \approx 3.1$--$3.4$), with $62$--$69\%$ of infeasible mass at $d\ge 3$ and support extending to the maximum distance $n(k-1)=10$. For reference, uniformly random bitstrings on this encoding yield an expected distance of $3.75$, so at $p=1$ under noise the penalized $\qubo$ cost Hamiltonian only weakly imprints the constraint structure on the sampled distribution, and the model relies primarily on post-selection. Importantly, $\mathbb{E}[d \mid d>0]$ is essentially constant in $t$ for every model (Figure~\ref{fig:violation_distance}, bottom): within the tight-to-naive range, the choice of penalty coefficient affects the frequency of infeasible samples and the quality of feasible ones, but not the severity of the violations, which appears to be governed by the formulation, encoding, and initialization.

In short, infeasible samples from the Dicke-initialized and XY-mixer models lie almost exclusively one bit flip from feasibility, $\rqubo$ violations are bounded and concentrated at $d\le 2$, whereas $\qubo$ violations average three vertex-level constraint violations regardless of the penalty value. This suggests that the penalty magnitude primarily affects how often the algorithm leaves the feasible space, while the formulation and initialization primarily affect how far it travels from it.

\section{Resource requirements}
\label{app:resources}

\begin{figure}[htbp]
    \centering
    \includegraphics[width=0.99\textwidth]{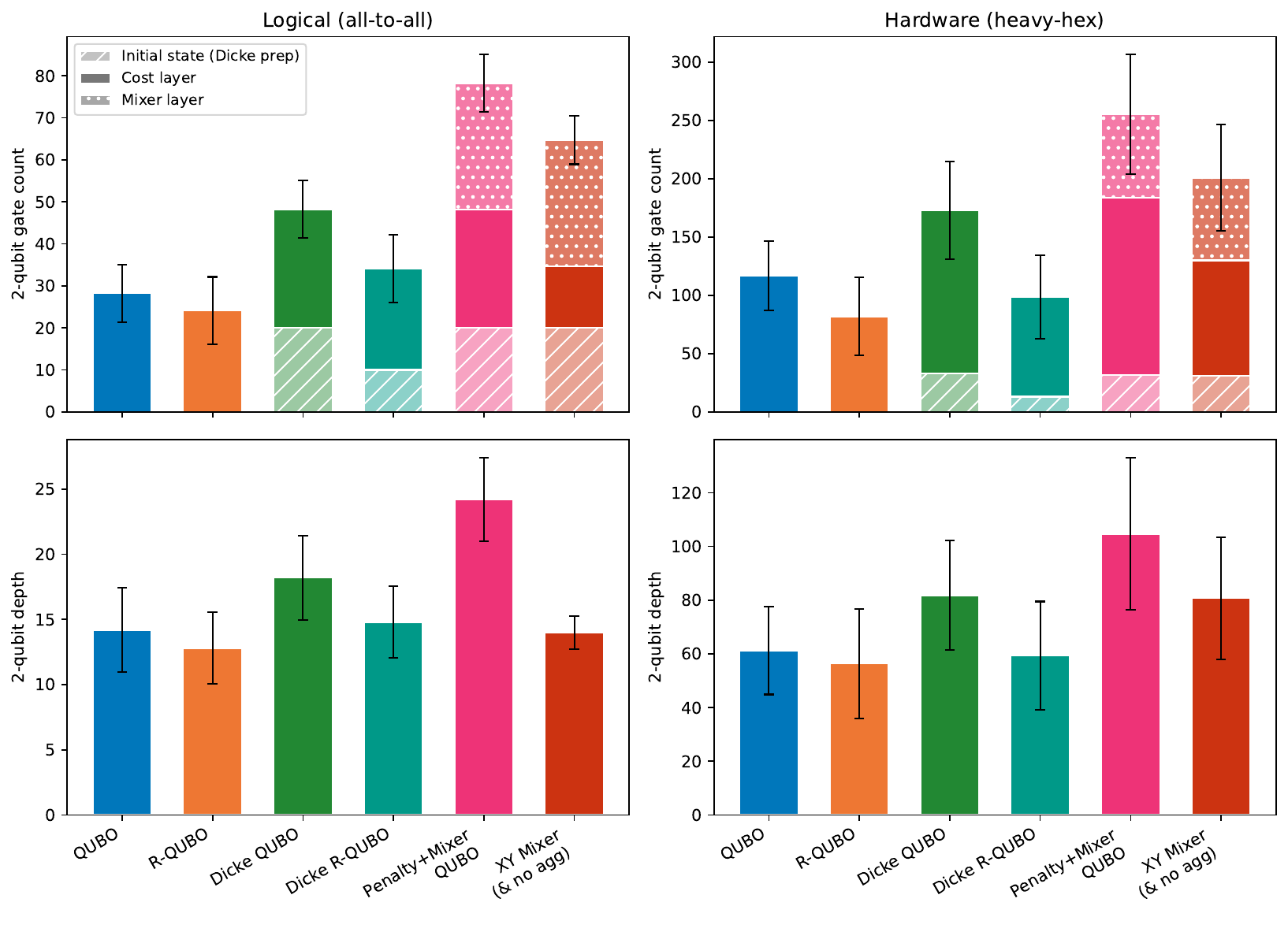}
    \caption{Two-qubit gate counts (top, stacked by circuit segment: initial state, cost layer, mixer layer) and two-qubit depth (bottom) for the six distinct QAOA circuits, at the logical level (left) and after transpilation to a 127-qubit heavy-hex {\tt IBM} backend (right). Bars show means over the 10 experiment graphs at $p=1$ at the tight-threshold setting; error bars show one standard deviation of the total. The XY-mixer models with and without CVaR aggregation share the same circuit.}
    \label{fig:resources}
\end{figure}

The solution quality and feasibility comparisons in Section~\ref{sec:num} should be weighed against the circuit cost of each formulation. Figure~\ref{fig:resources} reports two-qubit gate counts, separated by circuit segment (initial state, cost layer, mixer layer), and two-qubit depth for the six distinct circuits used in our experiments, both at the logical level and after transpilation to a 127-qubit heavy-hex {\tt IBM} backend.

The penalty-only formulations are the least expensive by a wide margin. The $\rqubo$ requires $n(k-1) = 10$ qubits and $81.9 \pm 33.2$ hardware two-qubit gates on average, versus $nk = 15$ qubits and $116.8 \pm 29.7$ gates for the $\qubo$; the reduced register lowers both the logical gate count and the routing overhead incurred on the heavy-hex topology. The Dicke $\rqubo$ adds only the reduced-Dicke preparation on top of the $\rqubo$ circuit: 10 logical ($13.0 \pm 3.0$ transpiled) two-qubit gates, versus 20 logical ($32.9 \pm 4.2$ transpiled) for one-hot Dicke preparation. It requires fewer quantum resources because the $k$-th partition set is encoded in the all-zeros string, so one branch of the superposition costs no additional entangling operations. Its total of $98.5 \pm 36.0$ hardware two-qubit gates is second only to the plain $\rqubo$. The constraint-preserving machinery is costly in comparison. Dicke state preparation adds 20 logical (approximately 33 hardware) two-qubit gates, and the XY-mixer adds 30 logical (approximately 71 hardware) two-qubit gates per QAOA layer. The XY-mixer cost layer is itself cheaper than the penalized one ($14.7$ versus $28.2$ logical two-qubit gates on average, since it carries no penalty terms), but the mixer more than offsets this saving: the full XY-mixer circuit requires $200.8 \pm 45.3$ hardware two-qubit gates, roughly $1.7\times$ the $\qubo$ and $2.5\times$ the $\rqubo$, and the Penalty+Mixer $\qubo$ is the most expensive at $255.1 \pm 51.3$. Moreover, the cost and mixer layers are repeated in each of the $p$ QAOA layers whereas state preparation is applied once, so the mixer overhead grows with depth.

\begin{figure}[htbp]
    \centering
    \includegraphics[width=0.9\textwidth]{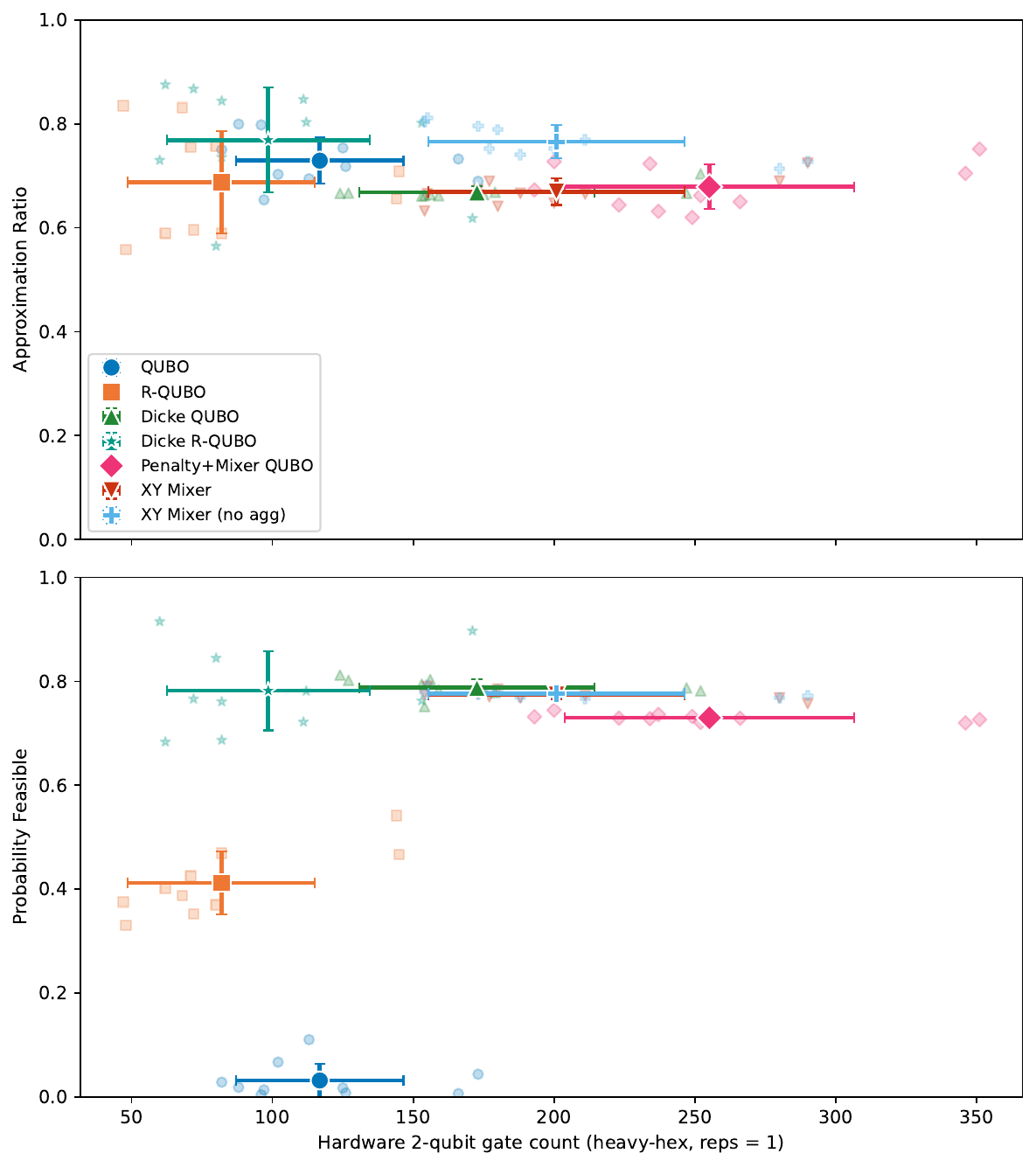}
    \caption{Solution quality versus circuit cost. Expected approximation ratio conditional on feasibility (top) and feasibility probability (bottom) at the tight-threshold setting, against each model's transpiled two-qubit gate count on its own graph. Faint markers: individual graphs; bold markers: mean $\pm$ one standard deviation. The Dicke $\rqubo$ occupies the upper-left (high quality, high feasibility, low cost) region of both panels.}
    \label{fig:resource_pareto}
\end{figure}

These counts quantify the trade-off discussed in the introduction: penalization adds terms to the cost Hamiltonian (here, approximately 14 logical two-qubit gates per layer for the one-hot penalties), while constraint-preserving mixers shift a larger cost into the mixer and circuit depth. They also help explain the noise-induced feasibility leakage of the XY-mixer observed in Section~\ref{sec:num}; the mixer layer adds roughly $71$ transpiled two-qubit gates per QAOA layer, raising the total circuit cost from $172.6$ (Dicke $\qubo$) to $255.1$ (Penalty+Mixer $\qubo$) two-qubit gates. Thus, the plain $\rqubo$ attains its solution quality with the smallest circuit of any model tested, while the Dicke $\rqubo$ delivers the highest feasible-sample quality and the highest feasibility at a hardware cost only ${\approx}20\%$ above it, roughly half the cost of the Dicke $\qubo$ and about $39\%$ of the Penalty+Mixer $\qubo$ gate count. Figure~\ref{fig:resource_pareto} summarizes these trade-offs by plotting solution quality and feasibility directly against circuit cost.

\end{document}